\documentclass[a4paper,11pt]{article}
\usepackage{jheppub} % for details on the use of the package, please see the JINST-author-manual
\usepackage{multirow}
\usepackage{lineno}
\usepackage{pgf}
\usepackage{soul}
\allowdisplaybreaks

\definecolor{ForestGreen}{RGB}{34, 139, 34} 

\arxivnumber{} % if you have one

\title{New strategies for probing $B\to D^\ast \ell\bar\nu_\ell$ lattice and experimental data}
\newcommand{\Fone}{{\mathcal{F}_1}}
\newcommand{\Ftwo}{{\mathcal{F}_2}}
\newcommand{\Vcb}{{|V_{cb}|}}
\newcommand{\BtoDstarellnu}{{B \to D^\ast\ell\bar\nu_\ell}}

\author[a]{M. Bordone,}
\author[a,b,c]{A. Jüttner}
\preprint{CERN-TH-2024-083}

\affiliation[b]{School of Physics and Astronomy, University of Southampton, Southampton, SO17 1BJ, UK}
\affiliation[c]{STAG Research Centre, University of Southampton, Southampton, SO17 1BJ, UK}
\affiliation[a]{Theoretical Physics Department, CERN, Geneva, Switzerland}

\emailAdd{marzia.bordone@cern.ch, andreas.juttner@cern.ch}

\abstract{We present an analysis of the exclusive semileptonic decay $B\to D^\ast \ell\bar\nu_\ell$ based on the Belle and Belle II data made public in 2023, combined with recent lattice-QCD calculations of the hadronic transition form factors by FNAL/MILC, HPQCD and JLQCD. We also consider a new combination of the Belle and Belle II data sets by HFLAV.
The analysis is based on the form-factor parameterisation by Boyd-Grinstein-Lebed (BGL), using Bayesian and frequentist statistics, for which we discuss novel strategies. We compare the results of an analysis where the BGL parameterisation is fit only to the lattice data with those from a simultaneous fit to lattice and experiment, and discuss the resulting predictions for the CKM-matrix element $\Vcb$, as well as other phenomenological observables, such as $R^{\tau/\mu}(D^\ast)$. We find tensions when comparing  analyses based on different combinations of experimental or theoretical input,  requiring the introduction of a systematic error for some of our results.
}

\begin{document}
\maketitle
\newpage
\section{Introduction}
The study of exclusive semileptonic $\BtoDstarellnu$ decays has, over the past years, received increasing attention. On the one hand, this is due to new experimental data becoming available from Belle~\cite{Belle:2023bwv,hepdata.137767} and Belle II~\cite{Belle-II:2023okj,hepdata.145129}. The former supersedes earlier results in~\cite{Belle:2017rcc}, while the latter is the first Belle-II analysis of the kinematic distribution of the decay. On the other hand, a new generation of lattice-QCD calculations of the corresponding hadronic transition form factors has recently been completed (see discussions in~\cite{FlavourLatticeAveragingGroupFLAG:2021npn,FLAG:2024,Tsang:2023nay}) by FNAL/MILC 21~\cite{FermilabLattice:2021cdg}, HPQCD 23~\cite{Harrison:2023dzh} and JLQCD 23~\cite{Aoki:2023qpa}, and further results are expected.
{These results are very important since the $\BtoDstarellnu$ decay channel} is one the most important players in the so-called $\Vcb$ puzzle, namely the discrepancy between the inclusive and exclusive $\Vcb$ determinations (see ~\cite{FlavourLatticeAveragingGroupFLAG:2021npn,FLAG:2024,ParticleDataGroup:2022pth,HFLAV:2022pwe} for a recent overview), and in the search for Lepton Flavour Universality (LFU) violation. 
Until a few years ago, form-factor calculations from Lattice QCD (LQCD) were available only at the kinematic endpoint $q^2_\mathrm{max} = (M_B-M_{D^*})^2$~\cite{FermilabLattice:2014ysv,Harrison:2017fmw}, with $q^2$ being the invariant mass of the lepton-neutrino pair. This required using experimental data, the Heavy Quark Effective Theory (HQET)~\cite{Falk:1992wt,Neubert:1993mb}, and calculations from Light-Cone Sum Rules \cite{Faller:2008tr,Bernlochner:2017jka,Jaiswal:2017rve,Gubernari:2018wyi,Gambino:2019sif,Bernlochner:2022ywh} to provide information complementary to the lattice results. 
The recent release of the above three lattice results provides form-factor information over a wider kinematic range for the first time. It is then possible to make predictions based entirely on first-principles theory, without the aid of experimental data or help from effective-field-theory or model calculations.

The scope of this work is to scrutinise the new LQCD and experimental results, and to make predictions for $\Vcb$ and other phenomenologically relevant observables, such as the LFU ratio $R^{\tau/\mu}(D^\ast)$. This requires further developing and testing analysis techniques. Here, we build on the analysis strategy recently developed in~\cite{Flynn:2023qmi,Flynn:2023eok} and then first applied to the decay $B_s\to K\ell\bar\nu_\ell$~\cite{Flynn:2023nhi}, which uses the combined power of Bayesian and frequentist statistics. We note that a similar effort based on the dispersive-matrix method~\cite{Lellouch:1995yv,BOURRELY1981157,DiCarlo:2021dzg,Martinelli:2021onb,Martinelli:2021myh,Martinelli:2022xir,Martinelli:2023fwm,Fedele:2023ewe} and analysing the data from the same collaborations has recently been accomplished in~\cite{Martinelli:2023fwm}. 

Our strategy is based on the model-independent Boyd-Grinstein-Lebed (BGL)~\cite{Boyd:1994tt} parameterisation of hadronic form factors. This allows testing the compatibility of the data with the Standard Model (SM) expectation, as well as predicting observables without residual truncation error, while taking constraints from quantum field theory, such as unitarity, consistently into account. Regarding the latter point, we note similar efforts based on the dispersive-matrix method~\cite{Lellouch:1995yv,BOURRELY1981157,Martinelli:2021onb,Martinelli:2021myh,Martinelli:2022xir,Martinelli:2023fwm,Fedele:2023ewe,DiCarlo:2021dzg} {or in a frequentist approach \cite{Bigi:2017njr,Bigi:2017jbd,Gambino:2019sif}}. We consider two different analysis strategies: first, parameterising the lattice-form-factor data and then combining with experimental information (similar to  e.g. LHCb~\cite{LHCb:2020ist} for the case of $B_s\to K\ell\bar\nu_\ell$, we will refer to this strategy as ``lat''), and second, simultaneously parameterising the lattice and experimental data (``lat+exp''). This requires extending the ideas of~\cite{Flynn:2023qmi}. We propose a novel procedure for the determination of $\Vcb$ in the first strategy based on the Akaike information criterion (AIC)~\cite{1100705,gamage2016adjusted}, which reduces a systematic effect discussed 
in~\cite{Gambino:2019sif,Ferlewicz:2020lxm}, that potentially originates from the d'Agostini bias~\cite{DAgostini:1993arp}.
Comparing the results of both the ``lat'' and ``lat+exp'' analyses allows for testing the SM in a  comprehensive way. Indeed, similar to~\cite{Martinelli:2021onb,Martinelli:2021myh,Martinelli:2023fwm,Fedele:2023ewe,Martinelli:2022xir}, we sometimes observe that theory predictions show unexpected behaviour, and also, that  results based on different experimental data  in some cases lead to conclusions that are at tension. We analyse how this affects the phenomenological predictions, and where deemed necessary, attach a corresponding systematic error.  

In what follows we first summarise the SM expression for the differential decay rate of $\BtoDstarellnu$ decays, as well as the BGL ansatz. We then discuss the two fitting strategies and results for the BGL parameterisations in Sec.~\ref{sec:The fitting problems} and~\ref{sec:Fit results}, respectively. In the remaining two sections we discuss the results for phenomenology and our conclusions.
\section{Anatomy of $\BtoDstarellnu$ decays}
We briefly introduce the expression for the differential decay rate for the process $\BtoDstarellnu$ in terms of hadronic form factors. Following that we discuss the model-independent parameterisation of the form factors, which are at the core of this study.
\subsection{Differential decay rates and hadronic form factors}
The semileptonic $\BtoDstarellnu$ decay, with the subsequent $D^*\to D\pi$ decay, is described by four kinematic variables. First is $q^2$, 
the square of the four-momentum transfer $q_\mu=(p_B-p_{D^\ast})_\mu$, where $p_B$ and $p_{D^*}$ are the four-momentum of the $B$ and the $D^\ast$ meson, respectively, or  equivalently the hadronic recoil
\begin{equation}
w=\frac{M_B^2+M_{D^\ast}^2-q^2}{2M_B M_{D^\ast}}\,.
\end{equation}
Second, there are three angles $\theta_\ell$, $\theta_v$ and $\chi$ that describe the geometry of the decay.\footnote{Following~\cite{Belle:2018ezy}, $\theta_\ell$ is the angle between the direction of movement of the charged lepton and the direction opposite the movement of the $B$ meson in the $W$ rest frame, $\theta_v$ is the angle between the direction of movement of the $D^0$ in the $D^0-\pi$ pair resulting from the decay of the $D^\ast$, and the direction opposite to the $B$ meson in the $D^\ast$ rest frame. The angle $\chi$ is the angle between the two decay planes defined by the charge-neutral lepton pair and the $D^0-\pi$ pair, respectively, in the $B$ rest frame. }
The expression for the differential decay rate in the SM in the limit of massless leptons  in terms of these kinematic variables is
\begin{align}\label{eq:diff dec rate}
    \frac{d\Gamma}{dwd\!\cos(\theta_\ell) d\!\cos(\theta_v)d\chi}=&
        \frac {3G_F^2}{1024\pi^4}|V_{cb}|^2\eta_{EW}^2M_B r^2\sqrt{w^2-1}q^2\nonumber\\
        &\hspace{-15mm}\times\big\{(1-\cos(\theta_\ell))^2\sin^2(\theta_v)H_+^2(w)
                    +(1+\cos(\theta_\ell))^2\sin^2(\theta_v)H_-^2(w)\nonumber\\
        &\hspace{-15mm}+4\sin^2(\theta_\ell)\cos^2(\theta_v)H_0^2(w)
                    -2\sin^2(\theta_\ell)\sin^2(\theta_v)\cos(2\chi)H_+(w)H_-(w)\nonumber\\
        &\hspace{-15mm}           -4\sin  (\theta_\ell)(1-\cos(\theta_\ell))\sin(\theta_v)\cos(\theta_v)\cos(\chi)H_+(w)H_0(w)\nonumber\\
        &\hspace{-15mm}            +4\sin  (\theta_\ell)(1+\cos(\theta_\ell))\sin(\theta_v)\cos(\theta_v)\cos(\chi)H_-(w)H_0(w)\big\}\,,
\end{align}
where $H_0$, $H_\pm$  are the hadronic helicity form factors defined in QCD. For massive charged leptons, an additional form factor contributes, which we denote with $H_S$.
For the discussion that follows, it is  convenient to use  also an alternative parameterisation of the functions $H_0$, $H_\pm$ and $H_S$ in terms of a new set of form factors $f$, $\Fone$, $\Ftwo$ and $g$, defined  as
\begin{align}\label{eq:helicity ff definition}
H_+(w)&=f(w)-M_B^2 r \sqrt{w^2-1}g(w)\,, &H_0(w)&=\frac{1}{\sqrt{q^2}}\Fone(w)\,,\nonumber\\
\\[-5mm]
 H_-(w)&=f(w)+M_B^2 r \sqrt{w^2-1}g(w)\,, & H_S(w)&=M_B r \frac{\sqrt{w^2-1}}{1-2 r w + r^2}\Ftwo(w)\,,\nonumber
\end{align}
which are subject to the kinematic constraints
\begin{align}\label{eq:kinematic constraints}
\mathcal{F}_1(1)&=M_B(1-r)f(1)\,,\nonumber\\
\\[-5mm]
\mathcal{F}_2(w_{\rm max})&=\frac{1+r}{M_B^2 r(1-r)(w_{\rm max}+1)}\mathcal{F}_1(w_{\rm max})\,,    \nonumber
\end{align}
where
\begin{equation}
w_{\rm max}=\frac{M_B^2+M_D^\ast}{2M_BM_{D^\ast}}=\frac{1+r^2}{2r}\qquad {\rm and}\quad r=\frac{M_{D^\ast}}{M_B}\,.
\end{equation}
Note that the form factors $f$, $\Fone$, $\Ftwo$ and $g$ can be classified in terms of their spin-parity quantum numbers as $1^+, \, 1^+,\, 0^-,$ and $1^-$.
The value $w_{\rm max}\approx 1.5$ corresponds to vanishing momentum transfer $q^2=0$, while $w=1$ corresponds to zero recoil, i.e.  $q^2_{\rm max}=(M_B-M_{D^\ast})^2$.

Simulations of LQCD predict the SM expectation for the form factors $f$, $\Fone$, $\Ftwo$ and $g$, and results are typically given at a small number of reference-$w$ values. The task for the following sections is therefore to determine a model-independent parameterisation of these form factors, taking into account the above kinematic constraints.

\subsection{Form-factor parameterisation}
We employ the ansatz by Boyd, Grinstein and Lebed (BGL)~\cite{Boyd:1994tt}, which is based on unitarity and analytic properties of hadronic form factors. At its core is the mapping of the complex $q^2$ plane with a cut along the positive real axis above $q^2=(M_B+M_{D^\ast})^2$, onto the unit disc in the new kinematic variable
\begin{equation}
z(w)\mapsto\frac{(\sqrt{w+1}-\sqrt{2})}{(\sqrt{w+1}+\sqrt{2})}\,.
\end{equation}
The values of $z$ for the physical semileptonic range are small ($z\in (0,0.056)$) and therefore particularly well suited for a polynomial expansion.
We now introduce two key elements of this parameterisation: 
The first one is the Blaschke factor $B_X$, which accounts for sub-threshold resonances, and is defined as
\begin{equation}
B_X(z)=\prod_k^{n_{\rm pole}}\frac{z-z_{{\rm pole},k}}{1-z z_{{\rm pole},k}}\,,\qquad
z_{{\rm pole},k}=\frac{\sqrt {t_+-M_{{X},k}^2}-\sqrt{t_+-t_-}}{\sqrt {t_+-M_{{X},k}^2}+\sqrt{t_+-t_-}}\,,
\end{equation}
with $t_\pm=(M_B\pm M_{D^\ast})^2$. The subscript $X$  refers to one of the form factors $f$, $\Fone$, $\Ftwo$ and $g$. The pole masses of the sub-threshold resonances $M_{X,k}$ are listed in Tab.~\ref{tab:inputs}.
The second element is the outer function $\phi_X$, defined as
\begin{equation}\label{eq:outer function}
        \phi_X(z) = N    \sqrt{\frac{n_I}{K\chi}}\
                  \frac{r^a(1+z)^a(1-z)^{b/2}}
                 {((1+r)(1-z)+2(\sqrt{r}(1+z)))^c}\,,
\end{equation}
where $\chi$ are the susceptibilities \cite{Bigi:2016mdz,Bigi:2017jbd,Bigi:2017njr,Martinelli:2021frl,Harrison:2024iad}, that are linked to vacuum-to-vacuum polarisation functions. Apart from $\chi$, $\phi_X(z)$ contains kinematic factors originating, amongst others, from the Jacobian of the variable change $w\to z(w)$. The coefficients  $N$, $n_I$, $K$, $a$, $b$ and $c$ and the susceptibilities are given in Tab.~\ref{tab:outer coefficients}. 
By construction, the product $B_X \phi_X F_X$ is analytic and can be Taylor expanded. Hence, each form factor can be parameterised as 
\begin{equation}
F_X(w)=\frac {1}{B_X(z)\phi_X(z)}\sum_{k=0}^{\infty}a_{X,k}(z(w))^k\,.
\label{eq:FF_param}
\end{equation}
The coefficients $a_{X,n}$ are \emph{a priori} not known and have to be determined from fits to LQCD and/or experimental data. However, the following unitarity  bounds on the BGL coefficients can be derived,
\begin{equation}\label{eq:unitarity constraints}
\sum_{k=0}^{\infty}|a_{f,k}|^2+|a_{\mathcal{F}_1,k}|^2\le1\,,\qquad
\sum_{k=0}^{\infty}|a_{g,k}|^2\le1\,,\qquad\sum_{k=0}^{\infty}|a_{\mathcal{F}_2,k}|^2\le1\,,\\\end{equation}
and imposed as part of the fitting procedure.
The bounds remain valid but are weakened after truncating the series in Eq.~(\ref{eq:FF_param}) at order $k =K_X$. Furthermore, the bounds together with the smallness of $z$ in the semileptonic range ensure good convergeance of the expansion. We implement the bounds following the strategy developed in~\cite{Flynn:2023qmi}, whereby unitarity is imposed as constraints on the likelihood integral within Bayesian inference when fitting form-factor parameterisations to lattice and/or experimental data. 
The kinematic constraints in Eq.~(\ref{eq:kinematic constraints})  are enforced by requiring
\begin{align}
a_{f,0}&=\frac{\phi_{f}(z(q^2_{\rm max})=0)}{\phi_{\mathcal{F}_1}(z(q^2_{\rm max})=0)}a_{\mathcal{F}_1,0}\qquad{\rm and}\label{eq:constraint1}\\
a_{\mathcal{F}_2,0}&=\frac{B_{\mathcal{F}_2}(z_{\rm max})\phi_{\mathcal{F}_2}(z_{\rm max})}{B_{\mathcal{F}_1}(z_{\rm max})\phi_{\mathcal{F}_1}(z_{\rm max})}\sum_{k=0}^{K_{\mathcal{F}_1}-1}a_{\mathcal{F}_1,k}\,z_{\rm max}^k-\sum_{k=1}^{K_{\mathcal{F}_2}-1}a_{\mathcal{F}_2,k}\,z_{\rm max}^k\,,\label{eq:constraint2}
\end{align}
 effectively eliminating the coefficients $a_{f,0}$ and $a_{\Ftwo,0}$.

\begin{table}
\begin{center}
\begin{tabular}{l|cccrrr lc}\hline\hline
         $X$&$N$&$n_I$&$K$&$a$&$b$&$c$& \multicolumn{1}{c}{$\chi$}\\\hline 
         $f$&     $4/M_B^2   $&2.6&$3\pi$&1& 3&4 & $3.894\cdot 10^{-4}\,\mathrm{GeV}^{-2}$ &\cite{Bigi:2016mdz,Bigi:2017jbd,Bigi:2017njr}\\
         $\Fone$ & $4/M_B^3$ & 2.6 &  $6\pi$ & 1 &  5 &  5 &$3.894\cdot 10^{-4}\,\mathrm{GeV}^{-2}$ &\cite{Bigi:2016mdz,Bigi:2017jbd,Bigi:2017njr}
        \\
         $\Ftwo$& $8\sqrt{2}$&2.6&$\pi $&2&-1&4 & $1.9421 \cdot 10^{-2}$ &\cite{Bigi:2016mdz,Bigi:2017jbd,Bigi:2017njr}\\
         $g$ &    $16$        &2.6&$3\pi$&2&-1&4 & $5.131\cdot 10^{-4}\,\mathrm{GeV}^{-2}$ &\cite{Bigi:2016mdz,Bigi:2017jbd,Bigi:2017njr}\\\hline\hline
\end{tabular}
\caption{Coefficients for the outer function defined in Eq.~(\ref{eq:outer function}).}
\label{tab:outer coefficients}
\end{center}
\end{table}

\section{The Fitting problems}\label{sec:The fitting problems}
With lattice data over a range of momentum-transfers~\cite{FermilabLattice:2021cdg,Harrison:2023dzh,Aoki:2023qpa} and also new experimental data~\cite{Belle:2023bwv,Belle-II:2023okj,hepdata.137767,hepdata.145129,HFLAV:2024} now available, we can consider  two fitting strategies. The first one relies on obtaining theory predictions for the form factors by fitting the BGL ansatz to LQCD data alone. These predictions, and their covariances are then combined with experimental data to obtain $\Vcb$. {We name this strategy ``lat''.} In the second strategy, we use the LQCD information together with
experimental data as fitting dataset, effectively extracting the BGL
parameters and $\Vcb$ at the same time.
{We call this second strategy ``lat+exp''.} In the following we discuss both strategies in detail.
\subsection{{Fit strategy one: ``lat'' only}}
\label{sec:BGL-fit to lattice data}
Input from computations of lattice QCD is typically given in terms of synthetic form-factor data at a small set of discrete kinematic reference points $w_{X,i}$. For the construction of the corresponding least-squares kernel we
collate input data and BGL parameters into vectors
\begin{equation}
{\bf f}^T=(
f_f(w_{f,0}),\dots,f_f(w_{f,N_f-1}),
f_{\Fone}(w_{{\mathcal{F}_1},0}),\dots,
f_{\Ftwo}(w_{{\mathcal{F}_2},0}),\dots,
f_{g}(w_{g,0}),\dots)\,,
\end{equation}
and
\begin{equation}
{\bf a}^T=
(a_{f,1},\dots,a_{f,K_f-1},\dots a_{\mathcal{F}_1,0},\dots,a_{\mathcal{F}_1,K_{\mathcal{F}_1}-1}, a_{\mathcal{F}_2,1},\dots,a_{\mathcal{F}_1,K_{\mathcal{F}_1}-1},
a_{g,0},\dots,a_{g,K_g-1})\,,\label{eq:avector}
\end{equation}
respectively, where we dropped the coefficients that are determined by the kinematic constraints
in Eqs. (\(\ref{eq:constraint1}\)) and (\(\ref{eq:constraint2}\)). 
The generalised linear least-squares kernel can now be written  as
\begin{equation}\label{eq:Chi_lat}
\chi^2_{\rm lat}({\bf a})=\left({\bf f}-Z\bf a\right)^TC_{\bf f}^{-1}\left({\bf f}-Z\bf a\right)\,,
\end{equation}
where $C_{\bf f}$ is the covariance matrix of the data ${\bf f}$. Due to its linear parameter dependence, the BGL expression for the form factor can be written as a matrix-vector product, ${\bf f}^{\rm BGL}=Z{\bf a}$, with the explicit form of the matrix $Z$ given in App.~\ref{app:A}.

We determine the parameters ${\bf a}$ using both Bayesian inference and frequentist fits. For the former, we follow~\cite{Flynn:2023qmi}, and impose the unitarity constraints~(\ref{eq:unitarity constraints}) in terms of a Bayesian prior. The constraint also acts as a regulator for higher-order terms and in principle allows us to increase the truncation $K_X$ arbitrarily. The frequentist fitting problem is solved by
\begin{equation}\label{eq:lin fit val}
{\bf a}= \left(Z^TC_{\bf f}^{-1}\right)^{-1}Z C_{\bf f}^{-1}{\bf f}\,,\qquad
    C_{\bf a}=\left(Z^TC^{-1}_{\rm f}Z\right)^{-1}\,,
\end{equation}
where $C_{\bf a}$ is the covariance matrix for the BGL coefficients. While the Bayesian-inference fit allows for a truncation-independent parameterisation of the data, the frequentist fit is limited by the number of degrees of freedom, but provides a measure for the quality of fit in terms of the $p$ value. We will make ample use of this complementarity.

In a second step, the resulting form-factor parameterisation can be combined with experimental input in order to compute the CKM-matrix element $|V_{cb}|$.
To this end, based on the BGL parameterisation, one first integrates the normalised differential decay rate Eq.~(\ref{eq:diff dec rate}) over phase space, restricting the integration with respect to $\alpha=w,\,\cos\theta_\ell,\,\cos\theta_v$ or $\chi$ to the range that corresponds to the experimental bin $i$, which in turn allows to compute 
\begin{equation}\label{eq:Vcb_bin}
	|V_{cb}|_{\alpha,i}=\left(
            \Gamma_{\rm exp}
            \left[\frac 1\Gamma\frac{d\Gamma}{d \alpha} \right]_{\rm exp}^{(i)}
            /
            \left[\frac{d\Gamma_0}{d \alpha}({\bf a} )\right]_{\rm lat}^{(i)}
            \right)^{1/2}\,,\qquad {\rm where}\qquad
            \Gamma_{\rm exp}=\frac{
             \mathcal{B}(B^0\to D^{\ast,-} \ell^+\nu_\ell)}
             {\tau(B^0)}\,,
\end{equation}
for each bin (see Tab.~\ref{tab:inputs} for experimental input), where we define $\Gamma_0= {\Gamma/|V_{cb}|^{2}}$.
A final result can then in principle be obtained as the result of a constant correlated fit over all results for $|V_{cb}|_{\alpha,i}$.
In practice however, we often find that such fits have acceptable $p$ values only after dropping bins, or, the fit result does not appear to represent the data well, an artefact that could be due to 
d'Agostini bias~\cite{DAgostini:1993arp}. Similar problems were also encountered in other studies~\cite{Gambino:2019sif,Ferlewicz:2020lxm,Martinelli:2021myh,Martinelli:2023fwm}. In order to mitigate these problems we propose to determine $|V_{cb}|$ in two alternative ways, namely,  by first computing correlated constant fits to all possible (in terms of fit quality, such that $0.05\le p\le 0.95$)
\begin{itemize}
    \item[a)] subsets $\{i\}$ of at least two bins in a given channel $\alpha$,
    \item[b)] subsets $\{\alpha,i\}$ of at least two bins chosen from any channel $\alpha$,
\end{itemize} 
and then combining them weighted by the Akaike-information criterion (AIC)~\cite{1100705,gamage2016adjusted}  (see \cite{Neil:2023pgt,Boyle:2022lsi,Jay:2020jkz,BMW:2014pzb,Borsanyi:2020mff} for other recent uses or discussions of the AIC). Contrary to the analysis in~\cite{Martinelli:2023fwm}, no PDG inflation~\cite{Workman:2022ynf} of the error at intermediate steps of the analysis is required and only \emph{good} fits enter the final result. The AIC weight factor for a given set $\{\alpha,i\}$ is
\begin{equation}\label{eq:AIC weight}
w_{\{\alpha,i\}} = \mathcal{N}^{-1}\exp\left(-\frac 12(\chi^2_{\{\alpha,i\}} - 2N_{{\rm dof},\{\alpha,i\}})\right)\,,\qquad{\rm where}\quad
\mathcal{N}=\sum\limits_{{\rm sets}\,\{\alpha,i\}}w_{{\rm set}}\,,
\end{equation}
where $\chi^2_{\{\alpha,i\}}$ is the correlated least-squares sum of the constant fit over $|V_{cb}|$ results for set $\{\alpha,i\}$, and 
$N_{{\rm dof},\{\alpha,i\}}$ is the corresponding number of degrees of freedom. The central value and error are then given as
\begin{equation}\label{eq:AIC mean}
    |V_{cb}|=\langle |V_{cb}|\rangle\equiv\sum
    \limits_{{\rm sets}\,\{\alpha,\,i\}}
    w_{\rm set}
    |V_{cb}|_{{\rm set}}
\end{equation}
and
\begin{equation}\label{eq:AIC error}
    \delta|V_{cb}|=\left(
    \langle \delta |V_{cb}|^2\rangle
    +\langle  |V_{cb}|^2\rangle
    -\langle  |V_{cb}|\rangle^2\right)^{1/2}
    \,,
\end{equation}
respectively. The first term under the square-root corresponds to a \emph{systematic} error from the variation of results under the AIC averaging, while the remaining terms correspond to the standard expression for the variance.

\subsection{{Fit strategy two: ``lat+exp''}}
Contrary to the strategy of the previous section, the simultaneous fit imposes the SM shape on the experimental data as well as unitarity bounds. The ``lat+exp'' strategy discussed in this section is, however, still interesting, since it provides complementary information for the search for NP. For sufficiently high precision of the lattice and experimental data, the presence of NP effects should lead to inconsistencies, or bad quality of fit. Alternatively, if NP effects are small compared to the statistical resolution of experiment and lattice, there could be enough freedom to account for small shifts in the fit results for the BGL parameters.
It is precisely such small modifications or inconsistencies that we are after in precision tests of the SM. Therefore, if the results of  simultaneous analyses differed from the ones based on the lattice-only fit, this could point to an interesting physics effect yet to be understood.

The simultaneous fit is defined in terms of the least-squares function
\begin{equation}\label{eq:Chi_combined}
\chi^2({\bf a})=\chi^2_{\rm lat}({\bf a}) + \chi^2_{\rm exp}({\bf a}) + \chi^2_{\rm norm}({\bf a},|V_{cb}|)\,.
\end{equation}
The first term is the contribution from the fit to lattice data defined in Eq.~(\ref{eq:Chi_lat}). The second term is the contribution from the binned normalised differential decay rate
\begin{equation}
    \chi_{\rm exp}^2({\bf a})=
    \sum\limits_{\alpha,i;\beta,j}
    \Delta_{\alpha,i}
    \left[C^{-1}_{ \Gamma}\right]_{\alpha,i;\beta,j}
    \Delta_{\beta,j}\,,\;\;{\rm where}\;\;
    \Delta_{\alpha,i}=\left[\frac 1\Gamma\frac{d\Gamma}{d \alpha} \right]_{\rm exp}^{(i)}- \left[\frac 1{\Gamma_0({\bf a})}\frac{d\Gamma_0}{d \alpha}({\bf a}) \right]_{\rm lat}^{(i)}\,,
\end{equation}
where $C_\Gamma$ is the covariance matrix of the normalised differential decay rate determined by experiment. The indices $\alpha$ and $\beta$ are summed over the set $\{w,\cos\theta_v,\cos\theta_\ell,\chi\}$, and $i$ and $j$ run over the  experimental bins in a given channel $\alpha$ or $\beta$.
The last term in Eq.~(\ref{eq:Chi_combined}) determines
 the overall normalisation
 \begin{equation}
 \chi^2_{\rm norm}({\bf a},|V_{cb}|) =\left(\Gamma_{\rm exp}-|V_{cb}|^2\Gamma_0({\bf a})\right)^2/\sigma^2_{\Gamma_{\rm exp}}\,,
 \end{equation}
 where $\Gamma_{\rm exp}$ is defined in Eq.~(\ref{eq:Vcb_bin}) with variance $\sigma^2_{\Gamma_{\rm exp}}$. This term is crucial for the determination of $|V_{cb}|$.

\section{Fit results}\label{sec:Fit results}
We now proceed with the discussion of the results for the BGL parameterisation following the two strategies laid out above. At each step, we discuss the complementary information gained from Bayesian inference and the frequentist fit. Details on the data curation of lattice and experimental data sets are given in App.~\ref{app:Lattice data sets} and~\ref{app:Experimental data sets}, respectively.
\begin{figure}
    \begin{center}
        \includegraphics[width=14cm]{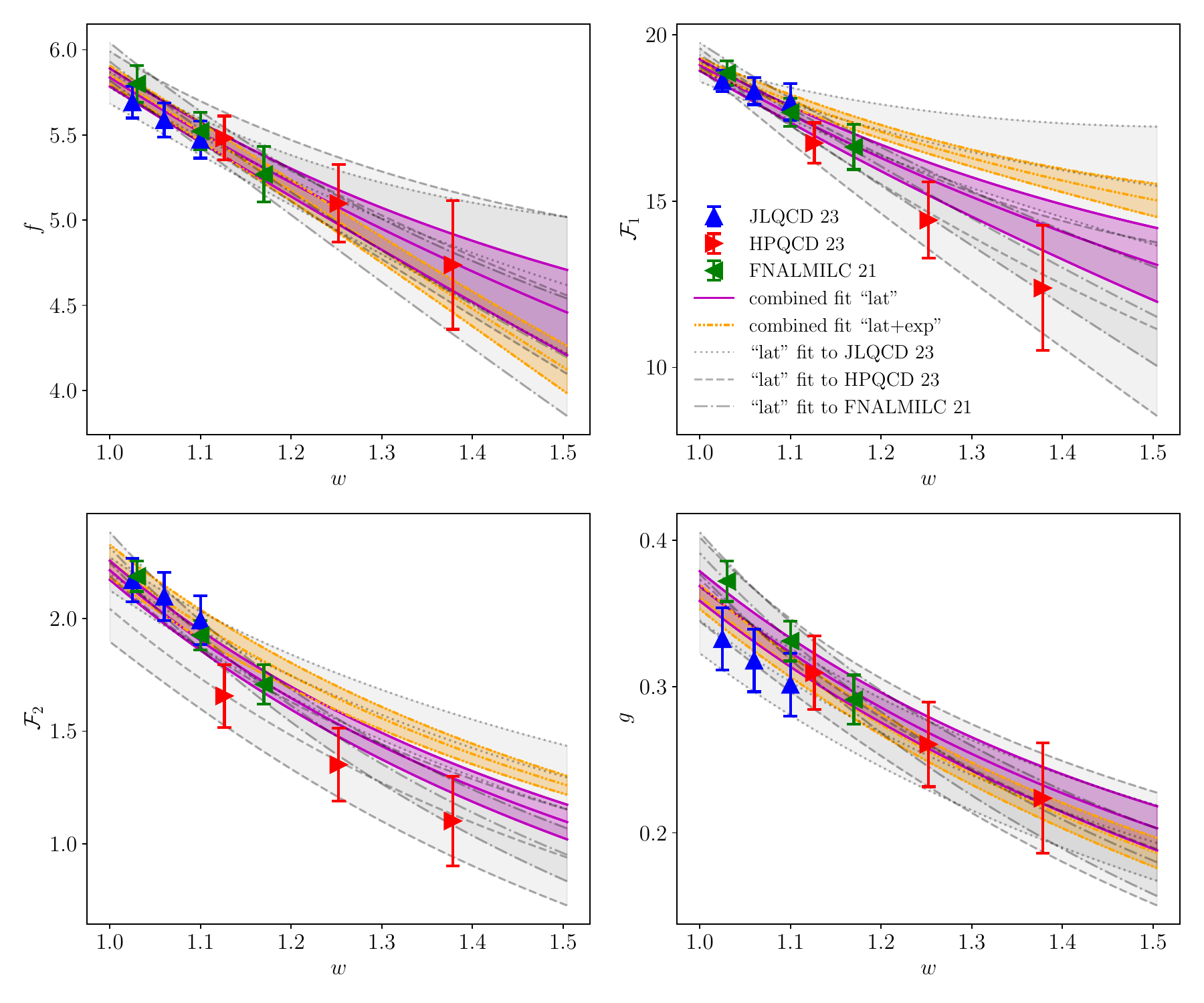}
    \end{center}
    \caption{{Solid magenta band: Simultaneous correlated Bayesian inference BGL ``lat'' fit to lattice results by JLQCD 23~\cite{Aoki:2023qpa}, HPQCD 23~\cite{Harrison:2023dzh} and FNAL/MILC 21~\cite{FermilabLattice:2021cdg} with $(K_f,K_\Fone,K_\Ftwo,g)=(4,4,4,4)$. Grey bands: Bayesian BGL ``lat'' fits to individual lattice data sets. Correlated frequentist fits to the same data without unitarity constraint would also be of acceptable quality:
    E.g. fit with $(K_f,K_\Fone,K_\Ftwo,K_g)=(2,2,2,2)$: $(p,\chi^2/N_{\rm dof},N_{\rm dof})=(0.95,0.62,30)$ or  $(K_f,K_\Fone,K_\Ftwo,g)=(4,4,4,4)$: $(p,\chi^2/N_{\rm dof},N_{\rm dof})=(0.79, 0.75, 22)$. The (densely dash dotted) orange band shows the corresponding simultaneous fit of JLQCD 23~\cite{Aoki:2023qpa}, HPQCD 23~\cite{Harrison:2023dzh} and FNAL/MILC 21~\cite{FermilabLattice:2021cdg} together with the experimental average of Belle 23~\cite{Belle:2023bwv,hepdata.137767} and Belle II 23~\cite{Belle-II:2023okj,hepdata.145129} in HFLAV 24~\cite{HFLAV:2024}, for which the frequentist fit would have $(p,\chi^2/N_{\rm dof},N_{\rm dof})=(0.13, 1.21, 58)$}.}\label{fig:BGL fit example}
\end{figure}
\subsection{BGL-fit to lattice data (``lat'')}\label{sec:lat fit}
\begin{figure}
    \begin{center}
        \includegraphics[width=14.5cm]{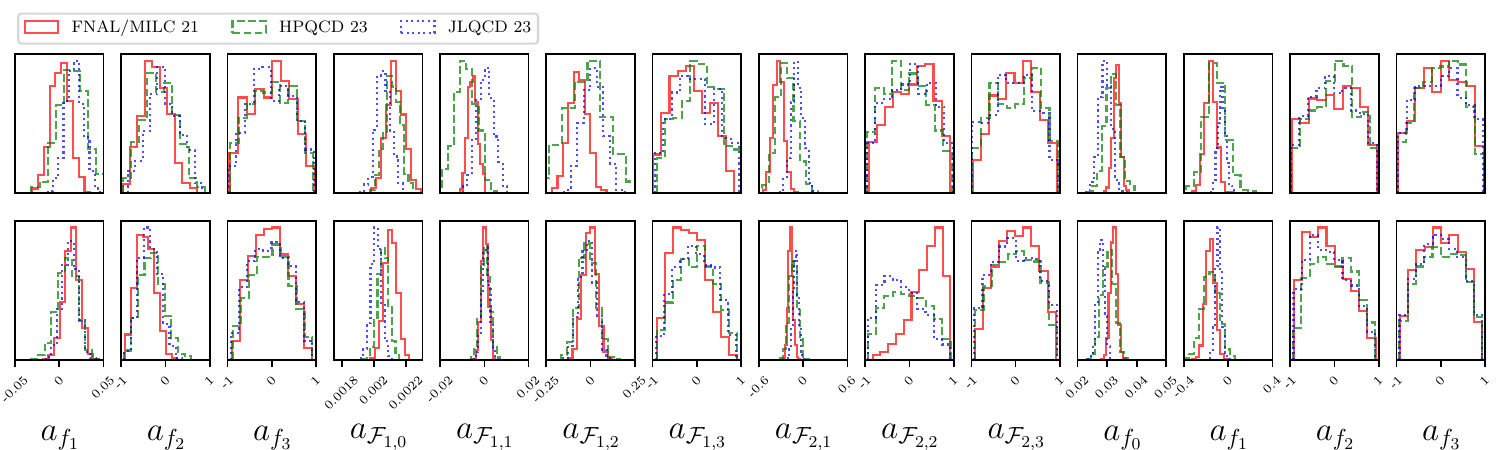}
    \end{center}
    \caption{Normalised histograms of the posterior distribution of BGL coefficients for the fits to individual lattice data (top, see Sec.~\ref{sec:lat fit}) and to the combined fit of these lattice data with experiment (bottom panel, see Sec.~\ref{sec:exp+lat fit}).}\label{fig:BGL histograms}
\end{figure}

A good overview over currently available lattice data, their compatibility, and the resulting BGL parameterisation can be gained from the plot in Fig.~\ref{fig:BGL fit example}. It shows a Bayesian-inference fit with $(K_f,K_\Fone,K_\Ftwo,K_g)=(4,4,4,4)$ to all the three lattice data sets FNAL/MILC 21, HPQCD 23 and JLQCD 23. The corresponding fit parameters and their stability as the truncation $K_X$ is increased from two to four for the Bayesian and frequentist fit can be seen in Tabs.~\ref{tab:Bayesian fit example} and~\ref{tab:Frequentist fit example}, respectively. As expected, the results for the first few significantly determined coefficients agree between both approaches. The higher-order coefficients 
in the frequentist fit, which are not constrained by the data, can assume values $\gg 1$ with large statistical errors. In Fig.~\ref{fig:BGL histograms} we show the distribution of BGL coefficients computed within Bayesian inference. The higher-order coefficients for $f, \Fone, \Ftwo$ and $g$, respectively, are not well determined by the data but constrained by the unitarity constraints~Eq.~(\ref{eq:unitarity constraints}). These ensure that the coefficients are smaller than unity, while no longer following a Gaussian distribution. The attached error therefore has to be interpreted with care.
Another aspect worth highlighting is the excellent frequentist fit quality indicated by the $p$ value and $\chi^2/N_{\rm dof}$ in the last three columns of Tab.~\ref{tab:Frequentist fit example}. 
In Tab.~\ref{tab:Bayesian fit variations} we also show  the BGL coefficients for the Bayesian BGL fit to individual or combinations of lattice-data sets.
We note differences in the 0th and 1st order BGL coefficients at the level of a few standard deviations. This is particularly the case for $a_{\Ftwo,1}$, where the tension between JLQCD 23 and FNAL/MILC 21 is about $2.5\sigma$, with HPQCD 23 lying in between for this coefficient. Consequently, the fit including JLQCD 23 and HPQCD 23 on the one side, and FNAL/MILC 21 and HPQCD 23 on the other, shows a similar tension. These deviations contribute to the different shapes of the  parameterisation of individual data sets as shown in terms of grey bands in Fig.~\ref{fig:BGL fit example}. We note, in this context, that FNAL/MILC 21 did not impose the kinematical constraint at maximum recoil in Eq.~(\ref{eq:constraint2}). Given that this constraint imposes a correlation between $\Fone$ and $\Ftwo$ at the level of the BGL coefficients, this might be a contributing element in this tension.

In summary, we find that all three LQCD data sets yield a good fit to the BGL parameterisation, even though individual results for some of the coefficients differ significantly. 
This observed quality of fit also holds for the simultaneous fit to all three lattice data sets. Note, however, that  statements about the quality of fits, or the compatibility of different data sets, are weakened by the fact that the  covariance matrix provided by the collaborations not only includes statistical, but also  systematic errors. This could lead to overly optimistic $p$ values. A clear separation of both effects, which would be required for a more reliable assessment, is unfortunately not possible with the available information. For the future it would therefore be desireable that lattice collaborations provide separate covariances for statistical and systematic effects, like for instance done in the case of $B_s\to K\ell\bar\nu_\ell$ in~\cite{Flynn:2023nhi}.
In terms of the more qualitative findings about the shape of form-factor parameterisations that can be extracted from Fig.~\ref{fig:BGL fit example}, we agree with the analysis of~\cite{Martinelli:2021onb,Martinelli:2021myh,Martinelli:2023fwm}. The analysis presented here provides a   quantitative understanding of the observations in terms of frequentist compatibility of the fit function with the data and a detailed comparison of BGL-fit coefficients.  

\begin{table}
\begin{center}
\begin{tabular}{l@{\hspace{1mm}}l@{\hspace{1mm}}l@{\hspace{1mm}}llllllllllllllllllllllllllllllllllllllllllllllll}
\hline\hline
$K_f$&$K_{\Fone}$&$K_{\Ftwo}$&$K_{g}$&\multicolumn{1}{c}{$a_{f,0}$}&\multicolumn{1}{c}{$a_{f,1}$}&\multicolumn{1}{c}{$a_{f,2}$}&\multicolumn{1}{c}{$a_{f,3}$}&\\
\hline
2&2&2&2&0.01223(12)&0.0118(58)&- &-&\\
3&3&3&3&0.01221(12)&0.0136(63)&-0.16(24)&-&\\
4&4&4&4&0.01221(11)&0.0133(64)&-0.14(23)&-0.01(49)&\\
\hline\hline
$K_f$&$K_{\Fone}$&$K_{\Ftwo}$&$K_{g}$&\multicolumn{1}{c}{$a_{{\Fone},0}$}&\multicolumn{1}{c}{$a_{{\Fone},1}$}&\multicolumn{1}{c}{$a_{{\Fone},2}$}&\multicolumn{1}{c}{$a_{{\Fone},3}$}&\\
\hline
2&2&2&2&0.002049(19)&-0.0042(15)&- &-&\\
3&3&3&3&0.002046(19)&-0.0039(16)&-0.019(36)&-&\\
4&4&4&4&0.002047(19)&-0.0038(17)&-0.016(42)&-0.00(46)&\\
\hline\hline
$K_f$&$K_{\Fone}$&$K_{\Ftwo}$&$K_{g}$&\multicolumn{1}{c}{$a_{{\Ftwo},0}$}&\multicolumn{1}{c}{$a_{{\Ftwo},1}$}&\multicolumn{1}{c}{$a_{{\Ftwo},2}$}&\multicolumn{1}{c}{$a_{{\Ftwo},3}$}&\\
\hline
2&2&2&2&0.04903(95)&-0.187(30)&- &-&\\
3&3&3&3&0.04896(90)&-0.201(41)&-0.04(56)&-&\\
4&4&4&4&0.04906(94)&-0.199(39)&-0.02(47)&0.00(51)&\\
\hline\hline
$K_f$&$K_{\Fone}$&$K_{\Ftwo}$&$K_{g}$&\multicolumn{1}{c}{$a_{g,0}$}&\multicolumn{1}{c}{$a_{g,1}$}&\multicolumn{1}{c}{$a_{g,2}$}&\multicolumn{1}{c}{$a_{g,3}$}&\\
\hline
2&2&2&2&0.03133(80)&-0.058(25)&- &-&\\
3&3&3&3&0.03129(81)&-0.062(27)&-0.10(55)&-&\\
4&4&4&4&0.03134(86)&-0.061(25)&-0.10(50)&-0.04(49)&\\
\hline\hline\\
\end{tabular}

\caption{Results for the simultaneous correlated Bayesian BGL fit to JLQCD 23~\cite{Aoki:2023qpa}, HPQCD 23~\cite{Harrison:2023dzh} and FNAL/MILC~21~\cite{FermilabLattice:2021cdg}.}\label{tab:Bayesian fit example}
\end{center}
\end{table}

\begin{table}
\begin{center}
 \resizebox{1\textwidth}{!}{  
\begin{tabular}{l@{\hspace{1mm}}l@{\hspace{1mm}}l@{\hspace{1mm}}llllllllllllllllllllllllllllllllllllllllllllllll}
\hline\hline
$K_f$&$K_{\Fone}$&$K_{\Ftwo}$&$K_{g}$&\multicolumn{1}{c}{$a_{f,0}$}&\multicolumn{1}{c}{$a_{f,1}$}&\multicolumn{1}{c}{$a_{f,2}$}&\multicolumn{1}{c}{$a_{f,3}$}&$p$&$\chi^2/N_{\rm dof}$&$N_{\rm dof}$&\\
\hline
2&2&2&2&0.01223(11)&0.0120(60)&- &-&0.95&0.62&30&\\
3&3&3&3&0.01221(12)&0.0136(70)&-0.19(31)&-&0.90&0.67&26&\\
4&4&4&4&0.01221(12)&0.0136(89)&-0.19(50)&-0.3(7.6)&0.79&0.75&22&\\
\hline\hline
$K_f$&$K_{\Fone}$&$K_{\Ftwo}$&$K_{g}$&\multicolumn{1}{c}{$a_{{\Fone},0}$}&\multicolumn{1}{c}{$a_{{\Fone},1}$}&\multicolumn{1}{c}{$a_{{\Fone},2}$}&\multicolumn{1}{c}{$a_{{\Fone},3}$}&$p$&$\chi^2/N_{\rm dof}$&$N_{\rm dof}$&\\
\hline
2&2&2&2&0.002049(19)&-0.0041(16)&- &-&0.95&0.62&30&\\
3&3&3&3&0.002046(19)&-0.0038(17)&-0.021(63)&-&0.90&0.67&26&\\
4&4&4&4&0.002046(21)&-0.0038(20)&-0.02(11)&-0.2(2.3)&0.79&0.75&22&\\
\hline\hline
$K_f$&$K_{\Fone}$&$K_{\Ftwo}$&$K_{g}$&\multicolumn{1}{c}{$a_{{\Ftwo},0}$}&\multicolumn{1}{c}{$a_{{\Ftwo},1}$}&\multicolumn{1}{c}{$a_{{\Ftwo},2}$}&\multicolumn{1}{c}{$a_{{\Ftwo},3}$}&$p$&$\chi^2/N_{\rm dof}$&$N_{\rm dof}$&\\
\hline
2&2&2&2&0.04903(93)&-0.186(31)&- &-&0.95&0.62&30&\\
3&3&3&3&0.04904(94)&-0.200(43)&-0.1(1.3)&-&0.90&0.67&26&\\
4&4&4&4&0.04902(94)&-0.195(62)&-0.4(3.0)&0.4(22.8)&0.79&0.75&22&\\
\hline\hline
$K_f$&$K_{\Fone}$&$K_{\Ftwo}$&$K_{g}$&\multicolumn{1}{c}{$a_{g,0}$}&\multicolumn{1}{c}{$a_{g,1}$}&\multicolumn{1}{c}{$a_{g,2}$}&\multicolumn{1}{c}{$a_{g,3}$}&$p$&$\chi^2/N_{\rm dof}$&$N_{\rm dof}$&\\
\hline
2&2&2&2&0.03138(87)&-0.059(24)&- &-&0.95&0.62&30&\\
3&3&3&3&0.03131(87)&-0.046(36)&-1.2(1.8)&-&0.90&0.67&26&\\
4&4&4&4&0.03126(87)&-0.017(48)&-3.7(3.3)&49.9(53.6)&0.79&0.75&22&\\
\hline\hline\\
\end{tabular}

}
\caption{Results for the simultaneous correlated frequentist BGL fit to JLQCD 23~\cite{Aoki:2023qpa}, HPQCD 23~\cite{Harrison:2023dzh} and FNAL/MILC 21~\cite{FermilabLattice:2021cdg}.}\label{tab:Frequentist fit example}
\end{center}
\end{table}
\begin{table}
    \begin{center}
     \resizebox{1\textwidth}{!}{  \begin{tabular}{l@{\hspace{1mm}}l@{\hspace{1mm}}l@{\hspace{1mm}}l@{\hspace{1mm}}llllllllllllllllllllllllllllllllllllllllllllllll}
\hline\hline
combination&\multicolumn{1}{c}{$a_{f,0}$}&\multicolumn{1}{c}{$a_{f,1}$}&\multicolumn{1}{c}{$a_{f,2}$}&\multicolumn{1}{c}{$a_{f,3}$}&$p$&$\chi^2/N_{\rm dof}$&$N_{\rm dof}$&\\
\hline
JLQCD 23&0.01209(19)&0.0175(99)&-0.04(35)&-0.00(46)&-&-&-&\\
HPQCD 23&0.01233(21)&0.015(15)&-0.12(36)&-0.01(47)&-&-&-&\\
FNALMILC 21&0.01241(23)&0.001(11)&-0.23(30)&-0.02(46)&-&-&-&\\
JLQCD 23, HPQCD 23&0.01218(13)&0.0149(78)&-0.04(29)&-0.02(46)&0.90&0.49&10&\\
JLQCD 23, FNALMILC 21&0.01220(14)&0.0133(75)&-0.08(28)&-0.00(48)&0.25&1.25&10&\\
FNALMILC 21, HPQCD 23&0.01233(14)&0.0054(85)&-0.28(25)&0.00(46)&0.94&0.41&10&\\
JLQCD 23, HPQCD 23, FNALMILC 21&0.01221(11)&0.0133(64)&-0.14(23)&-0.01(49)&0.79&0.75&22&\\
\hline\hline
combination&\multicolumn{1}{c}{$a_{{\Fone},0}$}&\multicolumn{1}{c}{$a_{{\Fone},1}$}&\multicolumn{1}{c}{$a_{{\Fone},2}$}&\multicolumn{1}{c}{$a_{{\Fone},3}$}&$p$&$\chi^2/N_{\rm dof}$&$N_{\rm dof}$&\\
\hline
JLQCD 23&0.002026(33)&0.0005(37)&0.016(59)&-0.02(47)&-&-&-&\\
HPQCD 23&0.002066(35)&-0.0084(48)&-0.02(12)&0.05(45)&-&-&-&\\
FNALMILC 21&0.002080(39)&-0.0052(22)&-0.070(51)&-0.13(42)&-&-&-&\\
JLQCD 23, HPQCD 23&0.002040(21)&-0.0025(25)&-0.002(50)&0.08(48)&0.90&0.49&10&\\
JLQCD 23, FNALMILC 21&0.002044(24)&-0.0036(18)&-0.002(43)&-0.11(45)&0.25&1.25&10&\\
FNALMILC 21, HPQCD 23&0.002067(24)&-0.0051(19)&-0.070(48)&-0.01(45)&0.94&0.41&10&\\
JLQCD 23, HPQCD 23, FNALMILC 21&0.002047(19)&-0.0038(17)&-0.016(42)&-0.00(46)&0.79&0.75&22&\\
\hline\hline
combination&\multicolumn{1}{c}{$a_{{\Ftwo
},0}$}&\multicolumn{1}{c}{$a_{{\Ftwo},1}$}&\multicolumn{1}{c}{$a_{{\Ftwo},2}$}&\multicolumn{1}{c}{$a_{{\Ftwo},3}$}&$p$&$\chi^2/N_{\rm dof}$&$N_{\rm dof}$&\\
\hline
JLQCD 23&0.0487(16)&-0.074(76)&0.01(50)&-0.05(49)&-&-&-&\\
HPQCD 23&0.0453(32)&-0.23(13)&-0.06(48)&0.02(50)&-&-&-&\\
FNALMILC 21&0.0513(15)&-0.332(69)&0.05(46)&0.02(46)&-&-&-&\\
JLQCD 23, HPQCD 23&0.0483(14)&-0.135(53)&-0.06(48)&-0.01(48)&0.90&0.49&10&\\
JLQCD 23, FNALMILC 21&0.0492(11)&-0.193(46)&0.06(48)&0.01(48)&0.25&1.25&10&\\
FNALMILC 21, HPQCD 23&0.0502(12)&-0.306(57)&0.03(45)&-0.02(47)&0.94&0.41&10&\\
JLQCD 23, HPQCD 23, FNALMILC 21&0.04906(94)&-0.199(39)&-0.02(47)&0.00(51)&0.79&0.75&22&\\
\hline\hline
combination&\multicolumn{1}{c}{$a_{g,0}$}&\multicolumn{1}{c}{$a_{g,1}$}&\multicolumn{1}{c}{$a_{g,2}$}&\multicolumn{1}{c}{$a_{g,3}$}&$p$&$\chi^2/N_{\rm dof}$&$N_{\rm dof}$&\\
\hline
JLQCD 23&0.0293(19)&-0.055(36)&-0.02(50)&0.00(50)&-&-&-&\\
HPQCD 23&0.0317(24)&-0.110(95)&0.02(49)&0.01(48)&-&-&-&\\
FNALMILC 21&0.0333(12)&-0.157(51)&-0.01(51)&0.01(49)&-&-&-&\\
JLQCD 23, HPQCD 23&0.0300(14)&-0.057(31)&0.00(50)&0.05(51)&0.90&0.49&10&\\
JLQCD 23, FNALMILC 21&0.03140(94)&-0.058(28)&-0.02(49)&-0.02(50)&0.25&1.25&10&\\
FNALMILC 21, HPQCD 23&0.0327(11)&-0.143(43)&-0.02(49)&-0.03(49)&0.94&0.41&10&\\
JLQCD 23, HPQCD 23, FNALMILC 21&0.03134(86)&-0.061(25)&-0.10(50)&-0.04(49)&0.79&0.75&22&\\
\hline\hline\\
\end{tabular}
}
\caption{
Results for the Bayesian $(K_f,K_\Fone,K_\Ftwo,g)=(4,4,4,4)$ BGL fits to combinations of lattice-data sets (see first column). Where available ($N_{\rm dof}\ge 1$) the quality of the corresponding frequentist fit is given in the right-most three columns.}\label{tab:Bayesian fit variations}
    \end{center}
\end{table}
\subsection{BGL-fit to lattice and experimental data (``lat+exp'')}\label{sec:exp+lat fit}
We start the discussion based on the example of the fit to the lattice and experimental data sets FNAL/MILC 21, HPQCD 23, JLQCD 23 and Belle II 23. With the inclusion in the fit of the experimental decay rates, the dependence on the BGL parameters is no longer linear. We therefore implemented the fit using the Python package \texttt{PyMultiNest} \cite{Feroz:2007kg,Feroz:2008xx,Feroz:2013hea,Buchner:2014nha} to sample the parameter space in the Bayesian approach.
Tabs.~\ref{tab:Bayesian fit example lat+exp} and~\ref{tab:Bayesian fit variations lat+exp} summarise the results for the BGL coefficients. As for the fit to only lattice data in the previous section we find that the fit parameters from the Bayesian-inference fit have converged to stable values for $(K_f,K_\Fone,K_\Ftwo,K_g)=(4,4,4,4)$. Coefficients starting with $a_{X,3}$ and higher are compatible with zero and are regulated by the unitarity constraint. Tab.~\ref{tab:Bayesian fit variations lat+exp} shows that these conclusions also hold for other choices of lattice input, where in each case a frequentist fit would also achieve perfectly acceptable quality of fit. As in the previous section, one finds that some BGL coefficients do vary by a few standard deviations between the three lattice data sets, while keeping the experimental input fixed. The tension in the order $i=1$ coefficients for the fit including different lattice data sets is now reduced, while some tensions in particular in the BGL coefficients of  $g$ at order $i=0$ are  exacerbated. The combined fit of all three lattice data sets together with HFLAV 24 is illustrated by the orange, densely dash dotted band in Fig.~\ref{fig:BGL fit example}. The magenta and orange band for all four form factors are compatible near vanishing recoil ($w=1$), and in particular the form factor $g$ agrees very well in shape. For $f$, $\Fone$ and $\Ftwo$, the inclusion of the experimental data into the fit changes the shape significantly, such that the form factors are visibly at tension at larger $w$, where the lattice data is at the same time least constraining due to large statistical errors or essentially due to the absence of lattice data points. A similar behaviour was also observed in~\cite{Fedele:2023ewe}.  It might at first be surprising that the form factor $\Ftwo$ is modified by the addition of experimental data. This form factor is proportional to $H_S$ (cf. Eq.~(\ref{eq:helicity ff definition})), which only enters the expression for the differential decay rate for massive leptons. However, the kinematical constraint in Eq.~(\ref{eq:kinematic constraints}) relates it to $\Fone$, which is controlled by the experimental data in the limit of massless leptons.
The variation in BGL coefficients is smaller when varying the experimental input while keeping the lattice input fixed, as summarised in Tab.~\ref{tab:Bayesian fit exp variations lat+exp}. There, the coefficients $a_{\Fone,1}$ and $a_{g,0}$ exhibit the largest tension. Due to its smaller errors for the normalised differential decay rate, it is the Belle-II 23 data that dominates in the fit to the  HFLAV 24 data set, as can be seen in Tab.~\ref{tab:Bayesian fit exp variations lat+exp}. The results for the CKM matrix element $|V_{cb}|$ that can be determined from the simultaneous fit to lattice and experimental data following Eq.~(\ref{eq:Chi_combined}) will be discussed in Sec.~\ref{sec:Extracting Vcb}. 
\begin{table}
\begin{center}
\begin{tabular}{l@{\hspace{1mm}}l@{\hspace{1mm}}l@{\hspace{1mm}}llllllllllllllllllllllllllllllllllllllllllllllll}
\hline\hline
$K_f$&$K_{\Fone}$&$K_{\Ftwo}$&$K_{g}$&\multicolumn{1}{c}{$a_{f,0}$}&\multicolumn{1}{c}{$a_{f,1}$}&\multicolumn{1}{c}{$a_{f,2}$}&\multicolumn{1}{c}{$a_{f,3}$}&\\
\hline
2&2&2&2&0.01230(11)&0.0064(44)&- &-&\\
3&3&3&3&0.01225(12)&0.0172(60)&-0.52(17)&-&\\
4&4&4&4&0.01226(11)&0.0161(61)&-0.47(16)&-0.03(39)&\\
\hline\hline
$K_f$&$K_{\Fone}$&$K_{\Ftwo}$&$K_{g}$&\multicolumn{1}{c}{$a_{{\Fone},0}$}&\multicolumn{1}{c}{$a_{{\Fone},1}$}&\multicolumn{1}{c}{$a_{{\Fone},2}$}&\multicolumn{1}{c}{$a_{{\Fone},3}$}&\\
\hline
2&2&2&2&0.002061(18)&-0.00033(55)&- &-&\\
3&3&3&3&0.002053(19)&-0.0004(11)&0.005(21)&-&\\
4&4&4&4&0.002054(19)&-0.0004(12)&0.010(33)&-0.09(37)&\\
\hline\hline
$K_f$&$K_{\Fone}$&$K_{\Ftwo}$&$K_{g}$&\multicolumn{1}{c}{$a_{{\Ftwo},0}$}&\multicolumn{1}{c}{$a_{{\Ftwo},1}$}&\multicolumn{1}{c}{$a_{{\Ftwo},2}$}&\multicolumn{1}{c}{$a_{{\Ftwo},3}$}&\\
\hline
2&2&2&2&0.05031(85)&-0.123(17)&- &-&\\
3&3&3&3&0.04998(88)&-0.131(28)&0.28(43)&-&\\
4&4&4&4&0.04998(88)&-0.128(26)&0.22(39)&0.00(46)&\\
\hline\hline
$K_f$&$K_{\Fone}$&$K_{\Ftwo}$&$K_{g}$&\multicolumn{1}{c}{$a_{g,0}$}&\multicolumn{1}{c}{$a_{g,1}$}&\multicolumn{1}{c}{$a_{g,2}$}&\multicolumn{1}{c}{$a_{g,3}$}&\\
\hline
2&2&2&2&0.03018(76)&-0.101(21)&- &-&\\
3&3&3&3&0.03034(78)&-0.087(24)&-0.34(45)&-&\\
4&4&4&4&0.03035(77)&-0.089(23)&-0.27(41)&-0.04(45)&\\
\hline\hline\\
\end{tabular}

\caption{Results for the simultaneous correlated Bayesian BGL fit to JLQCD 23~\cite{Aoki:2023qpa}, HPQCD 23~\cite{Harrison:2023dzh}, FNAL/MILC 21~\cite{FermilabLattice:2021cdg} and Belle II 23~~\cite{Belle-II:2023okj}.}\label{tab:Bayesian fit example lat+exp}
\end{center}
\end{table}
\begin{table}
    \begin{center} 
    \resizebox{1\textwidth}{!}{  
\begin{tabular}{l@{\hspace{1mm}}l@{\hspace{1mm}}l@{\hspace{1mm}}l@{\hspace{1mm}}llllllllllllllllllllllllllllllllllllllllllllllll}
\hline\hline
combination&\multicolumn{1}{c}{$a_{f,0}$}&\multicolumn{1}{c}{$a_{f,1}$}&\multicolumn{1}{c}{$a_{f,2}$}&\multicolumn{1}{c}{$a_{f,3}$}&$p$&$\chi^2/N_{\rm dof}$&$N_{\rm dof}$&\\
\hline
JLQCD 23&0.01202(18)&0.0123(86)&-0.35(22)&-0.03(42)&0.23&1.17&32&\\
HPQCD 23&0.01228(20)&0.009(11)&-0.30(26)&-0.01(41)&0.10&1.34&32&\\
FNALMILC 21&0.01256(23)&0.0142(86)&-0.45(21)&-0.04(39)&0.22&1.18&32&\\
JLQCD 23, HPQCD 23&0.01215(13)&0.0138(73)&-0.40(19)&-0.05(42)&0.36&1.06&44&\\
JLQCD 23, FNALMILC 21&0.01225(14)&0.0166(64)&-0.48(17)&-0.03(39)&0.14&1.23&44&\\
FNALMILC 21, HPQCD 23&0.01239(15)&0.0149(71)&-0.46(18)&-0.03(40)&0.19&1.18&44&\\
JLQCD 23, HPQCD 23, FNALMILC 21&0.01226(11)&0.0161(61)&-0.47(16)&-0.03(39)&0.18&1.17&56&\\
\hline\hline
combination&\multicolumn{1}{c}{$a_{{\Fone},0}$}&\multicolumn{1}{c}{$a_{{\Fone},1}$}&\multicolumn{1}{c}{$a_{{\Fone},2}$}&\multicolumn{1}{c}{$a_{{\Fone},3}$}&$p$&$\chi^2/N_{\rm dof}$&$N_{\rm dof}$&\\
\hline
JLQCD 23&0.002015(30)&0.0007(17)&-0.019(44)&-0.02(42)&0.23&1.17&32&\\
HPQCD 23&0.002059(33)&0.0005(21)&-0.018(51)&-0.01(42)&0.10&1.34&32&\\
FNALMILC 21&0.002105(38)&0.0003(15)&-0.004(37)&-0.17(37)&0.22&1.18&32&\\
JLQCD 23, HPQCD 23&0.002037(22)&0.0002(16)&-0.008(42)&-0.01(41)&0.36&1.06&44&\\
JLQCD 23, FNALMILC 21&0.002052(24)&-0.0002(12)&0.008(34)&-0.12(38)&0.14&1.23&44&\\
FNALMILC 21, HPQCD 23&0.002076(25)&-0.0001(13)&0.002(35)&-0.14(37)&0.19&1.18&44&\\
JLQCD 23, HPQCD 23, FNALMILC 21&0.002054(19)&-0.0004(12)&0.010(33)&-0.09(37)&0.18&1.17&56&\\
\hline\hline
combination&\multicolumn{1}{c}{$a_{{\Ftwo},0}$}&\multicolumn{1}{c}{$a_{{\Ftwo},1}$}&\multicolumn{1}{c}{$a_{{\Ftwo},2}$}&\multicolumn{1}{c}{$a_{{\Ftwo},3}$}&$p$&$\chi^2/N_{\rm dof}$&$N_{\rm dof}$&\\
\hline
JLQCD 23&0.0484(15)&-0.100(33)&-0.16(43)&-0.01(48)&0.23&1.17&32&\\
HPQCD 23&0.0505(25)&-0.130(56)&-0.04(46)&-0.00(46)&0.10&1.34&32&\\
FNALMILC 21&0.0524(15)&-0.169(32)&0.39(35)&0.03(44)&0.22&1.18&32&\\
JLQCD 23, HPQCD 23&0.0492(12)&-0.102(31)&-0.17(42)&-0.01(47)&0.36&1.06&44&\\
JLQCD 23, FNALMILC 21&0.05002(99)&-0.127(27)&0.19(40)&0.02(48)&0.14&1.23&44&\\
FNALMILC 21, HPQCD 23&0.0513(12)&-0.160(29)&0.40(34)&0.04(43)&0.19&1.18&44&\\
JLQCD 23, HPQCD 23, FNALMILC 21&0.04998(88)&-0.128(26)&0.22(39)&0.00(46)&0.18&1.17&56&\\
\hline\hline
combination&\multicolumn{1}{c}{$a_{g,0}$}&\multicolumn{1}{c}{$a_{g,1}$}&\multicolumn{1}{c}{$a_{g,2}$}&\multicolumn{1}{c}{$a_{g,3}$}&$p$&$\chi^2/N_{\rm dof}$&$N_{\rm dof}$&\\
\hline
JLQCD 23&0.0279(12)&-0.086(27)&-0.12(46)&0.00(46)&0.23&1.17&32&\\
HPQCD 23&0.0303(21)&-0.159(74)&-0.02(46)&0.01(46)&0.10&1.34&32&\\
FNALMILC 21&0.0323(12)&-0.160(41)&-0.15(45)&0.00(45)&0.22&1.18&32&\\
JLQCD 23, HPQCD 23&0.02871(98)&-0.086(26)&-0.16(43)&0.00(47)&0.36&1.06&44&\\
JLQCD 23, FNALMILC 21&0.03026(84)&-0.088(24)&-0.27(42)&-0.01(45)&0.14&1.23&44&\\
FNALMILC 21, HPQCD 23&0.0318(10)&-0.153(37)&-0.14(44)&-0.01(46)&0.19&1.18&44&\\
JLQCD 23, HPQCD 23, FNALMILC 21&0.03035(77)&-0.089(23)&-0.27(41)&-0.04(45)&0.18&1.17&56&\\
\hline\hline\\
\end{tabular}
}
\caption{
Results for the Bayesian $(K_f,K_\Fone,K_\Ftwo,g)=(4,4,4,4)$ BGL fits to Belle II 23 and combinations of lattice-data sets (see first column). The quality of the corresponding frequentist fit is given in the right-most three columns.}\label{tab:Bayesian fit variations lat+exp}
    \end{center}
\end{table}
\begin{table}
    \begin{center} 
\begin{tabular}{l@{\hspace{1mm}}l@{\hspace{1mm}}l@{\hspace{1mm}}l@{\hspace{1mm}}llllllllllllllllllllllllllllllllllllllllllllllll}
\hline\hline
combination&\multicolumn{1}{c}{$a_{f,0}$}&\multicolumn{1}{c}{$a_{f,1}$}&\multicolumn{1}{c}{$a_{f,2}$}&\multicolumn{1}{c}{$a_{f,3}$}&$p$&$\chi^2/N_{\rm dof}$&$N_{\rm dof}$&\\
\hline
Belle 23&0.01223(11)&0.0153(60)&-0.30(19)&-0.02(44)&0.25&1.12&58&\\
BelleII 23&0.01226(11)&0.0161(61)&-0.47(16)&-0.03(39)&0.18&1.17&56&\\
HFLAV 23&0.01224(11)&0.0157(56)&-0.50(16)&-0.05(39)&0.13&1.21&58&\\
\hline\hline
combination&\multicolumn{1}{c}{$a_{{\Fone},0}$}&\multicolumn{1}{c}{$a_{{\Fone},1}$}&\multicolumn{1}{c}{$a_{{\Fone},2}$}&\multicolumn{1}{c}{$a_{{\Fone},3}$}&$p$&$\chi^2/N_{\rm dof}$&$N_{\rm dof}$&\\
\hline
Belle 23&0.002050(19)&-0.0023(13)&0.018(34)&0.11(41)&0.25&1.12&58&\\
BelleII 23&0.002054(19)&-0.0004(12)&0.010(33)&-0.09(37)&0.18&1.17&56&\\
HFLAV 23&0.002052(19)&-0.0008(11)&0.011(31)&0.00(37)&0.13&1.21&58&\\
\hline\hline
combination&\multicolumn{1}{c}{$a_{{\Ftwo},0}$}&\multicolumn{1}{c}{$a_{{\Ftwo},1}$}&\multicolumn{1}{c}{$a_{{\Ftwo},2}$}&\multicolumn{1}{c}{$a_{{\Ftwo},3}$}&$p$&$\chi^2/N_{\rm dof}$&$N_{\rm dof}$&\\
\hline
Belle 23&0.04958(89)&-0.148(26)&0.38(36)&0.03(45)&0.25&1.12&58&\\
BelleII 23&0.04998(88)&-0.128(26)&0.22(39)&0.00(46)&0.18&1.17&56&\\
HFLAV 23&0.04996(85)&-0.132(25)&0.26(38)&0.02(46)&0.13&1.21&58&\\
\hline\hline
combination&\multicolumn{1}{c}{$a_{g,0}$}&\multicolumn{1}{c}{$a_{g,1}$}&\multicolumn{1}{c}{$a_{g,2}$}&\multicolumn{1}{c}{$a_{g,3}$}&$p$&$\chi^2/N_{\rm dof}$&$N_{\rm dof}$&\\
\hline
Belle 23&0.03135(76)&-0.064(23)&-0.13(44)&-0.01(46)&0.25&1.12&58&\\
BelleII 23&0.03035(77)&-0.089(23)&-0.27(41)&-0.04(45)&0.18&1.17&56&\\
HFLAV 23&0.03072(72)&-0.082(22)&-0.26(42)&-0.01(46)&0.13&1.21&58&\\
\hline\hline\\
\end{tabular}

\caption{
Results for the Bayesian $(K_f,K_\Fone,K_\Ftwo,g)=(4,4,4,4)$ BGL fits to Belle 23, Belle II 23 or HFLAV 24, in each case jointly with the lattice data sets FNAL/MILC 21, HPQCD 23 and JLQCD 23. The quality of the corresponding frequentist fit is given in the right-most three columns.}\label{tab:Bayesian fit exp variations lat+exp}
    \end{center}
\end{table}
\subsection{Comparison of fit results}
\begin{figure}
    \begin{center}
        \includegraphics[width=12cm]{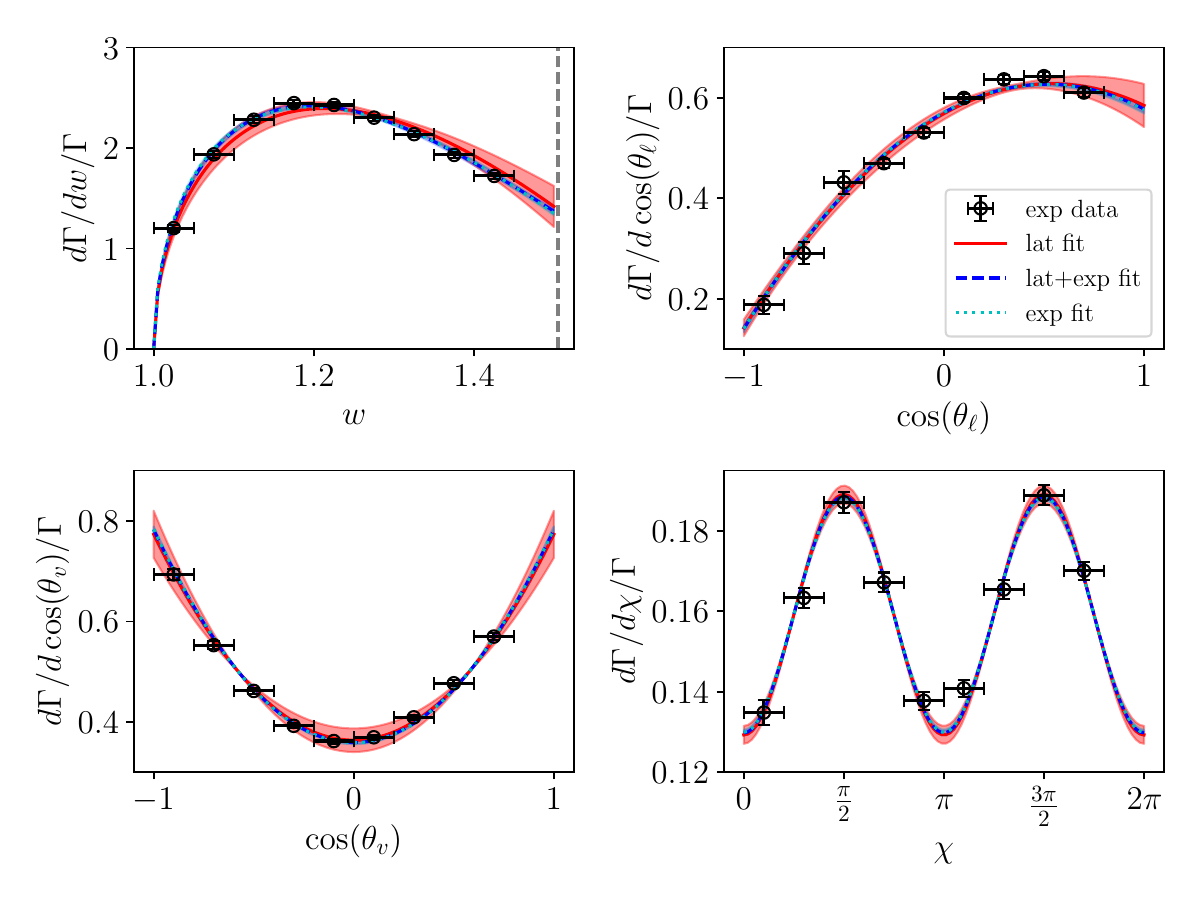}
        \includegraphics[width=12cm]{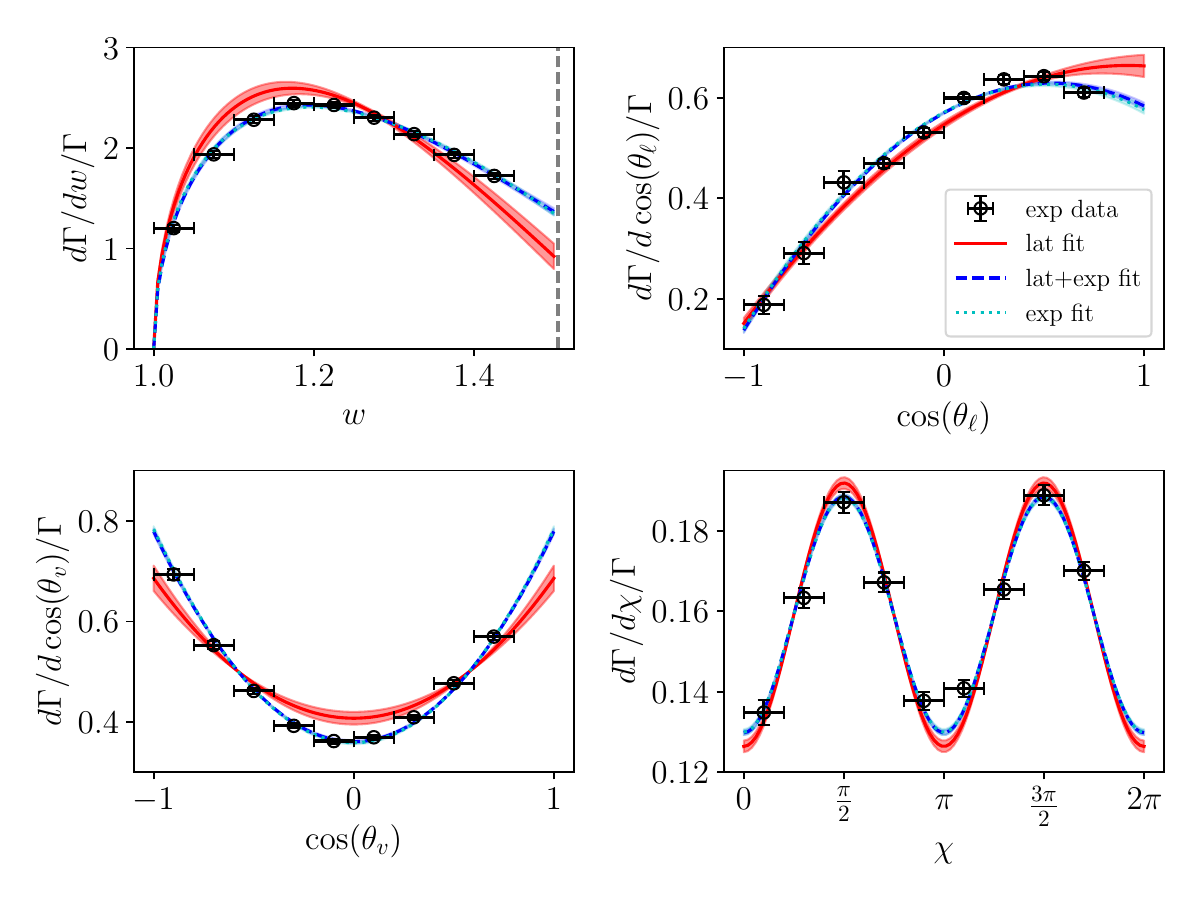}
    \end{center}
    \caption{Differential decay rates by HFLAV  24 and a $(K_f,K_\Fone,K_\Ftwo,g)=(4,4,4,4)$ BGL fits to lat, lat+exp and exp. Lattice input JLQCD 23 (top two rows) and FNAL/MILC 21 and HPQCD 23 (bottom two rows). We show the 1$\sigma$ error bands. The horizontal bars indicate the bin-width of the experimental data.}\label{fig:dGdX}
\end{figure}

Fig.~\ref{fig:dGdX} shows the HFLAV 24 differential decay rates together with the BGL fits to JLQCD 23 (top two rows) and FNAL/MILC 21 and HPQCD 23 (bottom two rows). While the lattice-only fit from Sec.~\ref{sec:lat fit} (red line and band) based on JLQCD 23 nicely agrees with the shapes of the differential decay rate, which is a highly non-trivial outcome, the result of the combined fit to FNAL/MILC 21 and HPQCD 23 appears to miss many experimental data points. The same is observed for the fits including only FNAL/MILC 21 or HPQCD 23, respectively. The combined lattice and experiment fits from Sec.~\ref{sec:exp+lat fit} (blue dashed line and band) in both cases nicely agree with the data points as expected by the good quality of fit observed in the previous section. Inspecting the BGL coefficients in Tabs.~\ref{tab:Bayesian fit variations} (``lat'') and~\ref{tab:Bayesian fit variations lat+exp} (``lat+exp'') we find that for both FNAL/MILC 21 and HPQCD 23, in particular the coefficients $a_{X,1}$ vary up to a few standard deviations  in order to accommodate the shape imposed by the experimental data. This shift does not deteriorate the quality of fit, i.e., the lattice data can accommodate this change in shape of the form factors, in particular for $\Fone$ and $\Ftwo$.  For JLQCD 23 there is less variation, i.e., the lattice data alone more naturally describes the shape of the differential decay rate found  in experiment. In the bottom panel of Fig~\ref{fig:BGL histograms} we show the posterior distributions of the BGL ``lat+exp'' fit to lattice and experimental data, where the observed shifts are also visible, comparing to the lattice-only ``lat'' fit in the top panel. We also note that the inclusion of experimental data in the case of FNAL/MILC~21 appears to \emph{pull} the result for $a_{\Ftwo,2}$ towards the upper limit of what is allowed by the unitarity constraint. This is not happening for HPQCD 23 and JLQCD 23, respectively. As stated above, FNAL/MILC~21 did not impose the kinematic constraint in Eq.~(\ref{eq:kinematic constraints}) that relates $\Fone$ and $\Ftwo$ in their form-factor parameterisation, and this might provide the key to the observed behaviour.
\begin{figure}
    \begin{center}
        \includegraphics[width=13cm]{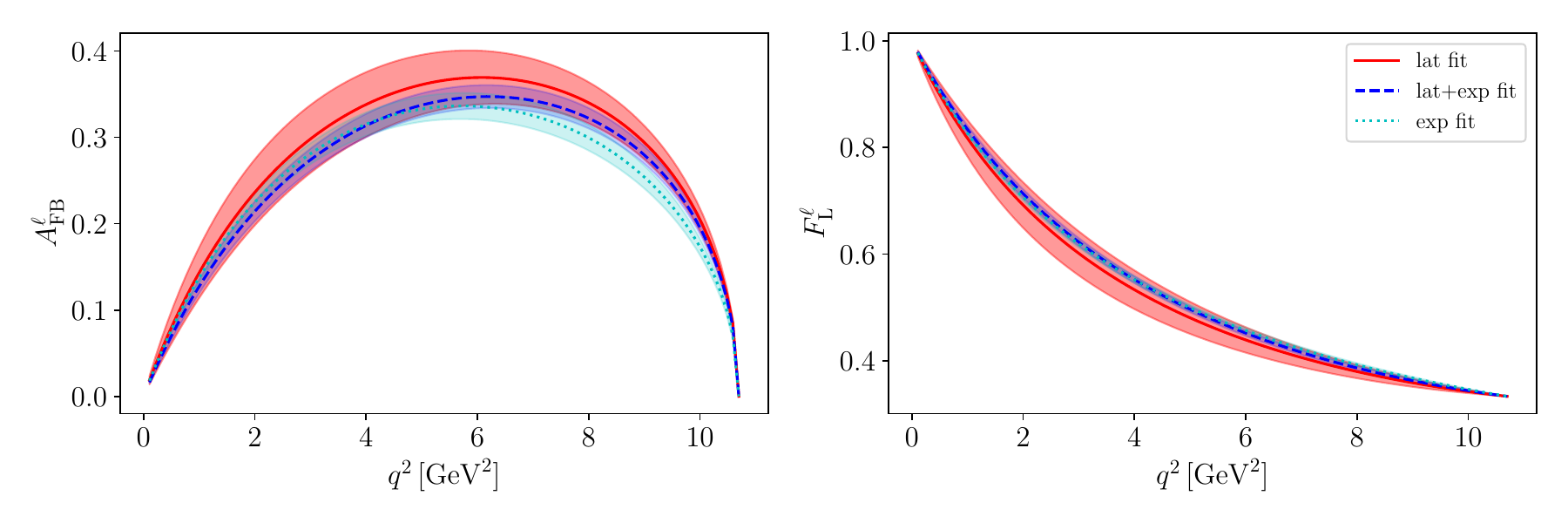}
        \includegraphics[width=13cm]{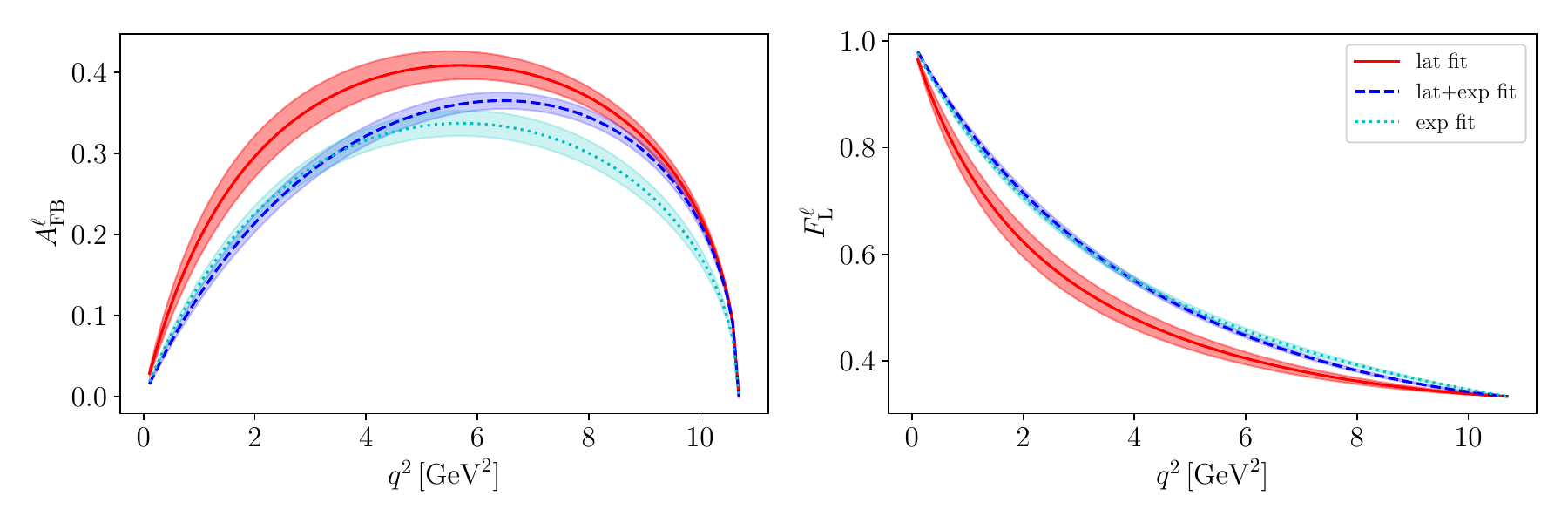}
    \end{center}
    \caption{Asymmetries $A_{\rm FB}^\ell$ and $F_{L}^\ell$ in the limit of massless charged leptons and based on  $(K_f,K_\Fone,K_\Ftwo,g)=(4,4,4,4)$ BGL fits to lat, lat+exp and exp, based on Belle II 23 for experiment and JLQCD 23 (top row), and HPQCD 23 and FNAL/MILC 21 (bottom row) for lattice.}\label{fig:angular observables}
\end{figure}

We can now have a first look at two angular observables for the decay $B\to D^{*}(
\to D\pi)\ell\bar\nu_\ell$. Introducing the normalisation
\begin{equation}
    \mathcal{N}_{m_\ell}=I_{m_\ell}\left[\left(1+\frac{m_\ell^2}{2q^2}\right)\left(H_0^2 + H_-^2 + H_+^2\right)+\frac 32\frac{ m_\ell^2}{q^2} H_S^2\right]\,,
\end{equation}
the forward-backward asymmetry is
\begin{equation}\label{eq:AFB}
     A^\ell_{\rm FB}=
     \frac{\int_{0}^1-\int_{-1}^0 d \cos\theta_\ell d\Gamma/d\cos\theta_\ell}{
     \int_{0}^1+\int_{-1}^0 d\cos\theta_\ell d\Gamma/d\cos\theta_\ell
     }=
     \frac 34 I_{m_\ell}\left[H_-^2-H_+^2-\frac{2m_\ell^2}{q^2}H_0 H_S\right]/\mathcal{N}_{m_\ell}\,,
\end{equation}
and the longitudinal $D^\ast$ polarisation fraction~\cite{Fajfer:2012vx}
\begin{equation}\label{eq:FL}
         {F^\ell_L= I_{m_\ell}\left[H_0^2\left(1+\frac{m_\ell^2}{2q^2}\right)+\frac{3 m_\ell^2}{2q^2}H_S^2\right]/\mathcal{N}_{m_\ell}\,,}
\end{equation}
where 
\begin{equation}
    I_{m_\ell}[f] = \frac{1}{M_B M_{D^\ast}}
        \int_{m_{\ell}^2}^{q^2_{\rm max}}\sqrt{w^2-1}\,q^2
        (1-m_\ell^2/q^2)^2f(q^2)dq^2\,.
    \label{eq:PS_integral}
\end{equation}
The plots in Fig.~\ref{fig:angular observables} show these ratios before phase-space integration in the numerator and denominator, respectively, in the case of massless  charged leptons in the final state. These plots are instructive, since they provide another illustration of the difference in shape of the lattice form factors. The plots show again the fit based on only JLQCD 23 (top row) on the one side, and FNAL/MILC 21 and HPQCD 23 (bottom row) on the other side. In the former case the shapes are largely compatible between the ``lat'', ``lat+exp'' and experiment-only fits. As already observed in~\cite{Fedele:2023ewe}, a clear and significant difference in the shapes can be observed in the latter  case. We will return to this tension below when discussing the integrated versions, i.e., $A^\ell_{\rm FB}$ and $F^\ell_L$, respectively, which allows for a more quantitative statement of this observation. 

To summarise, for the data sets at hand a number of tensions appear between lattice and experimental data sets in the analysis following the two strategies ``lat'' and ``lat+exp''. Some of these tensions have been observed before in~\cite{Martinelli:2023fwm, Fedele:2023ewe} based on the ``lat'' analysis within the dispersive-matrix method. Here we provide a complementary view in terms of the results of the ``lat+exp'' analysis of all three lattice data sets and the Belle 23 and Belle II 23 experimental data sets. The Bayesian-inference framework based on the BGL expansion used here, allows to relate the observations to tensions  in the  BGL coefficients. Within the  frequentist approach the tensions are however not sufficiently strong to allow to identify their origin in terms of a problem with either lattice or experimental data, or merely statistical fluctuations. 
Revisiting the situation in the future with new and hopefully more precise lattice and experimental data therefore remains an exciting outlook.

\section{Phenomenology}
The previous sections concentrated on the results for BGL fits to lattice and experimental data. In this section we discuss results with relevance for phenomenology and compare to the literature.
\subsection{Determination of $|V_{cb}|$}\label{sec:Extracting Vcb}
\subsubsection{$\Vcb$ from the ``lat'' fit}\label{sec:Vub lat}
\begin{figure}
    \begin{center}
    FNAL/MILC 21\\
        \includegraphics[width=12cm]{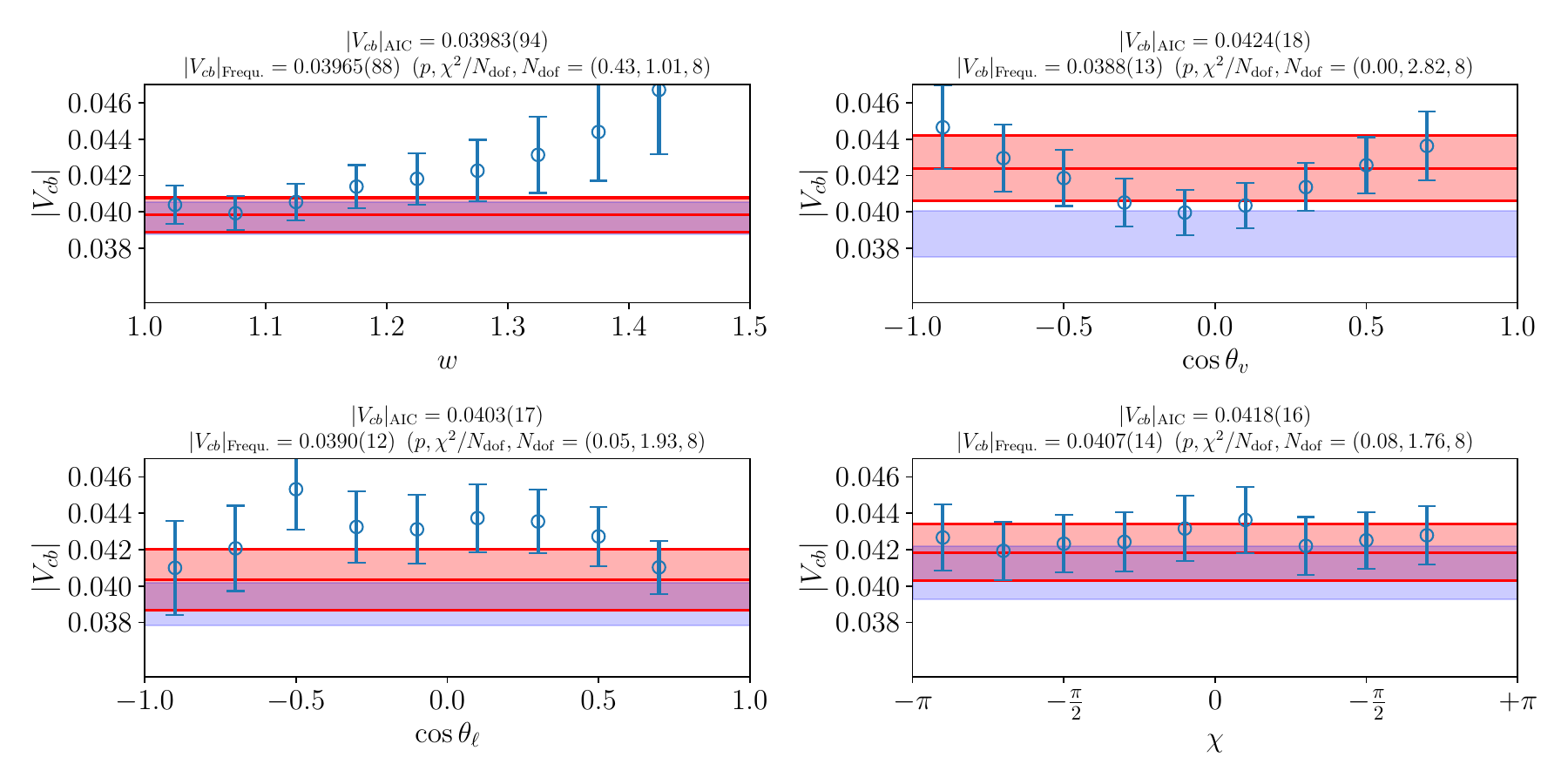}\\
        HPQCD 23\\
        \includegraphics[width=12cm]{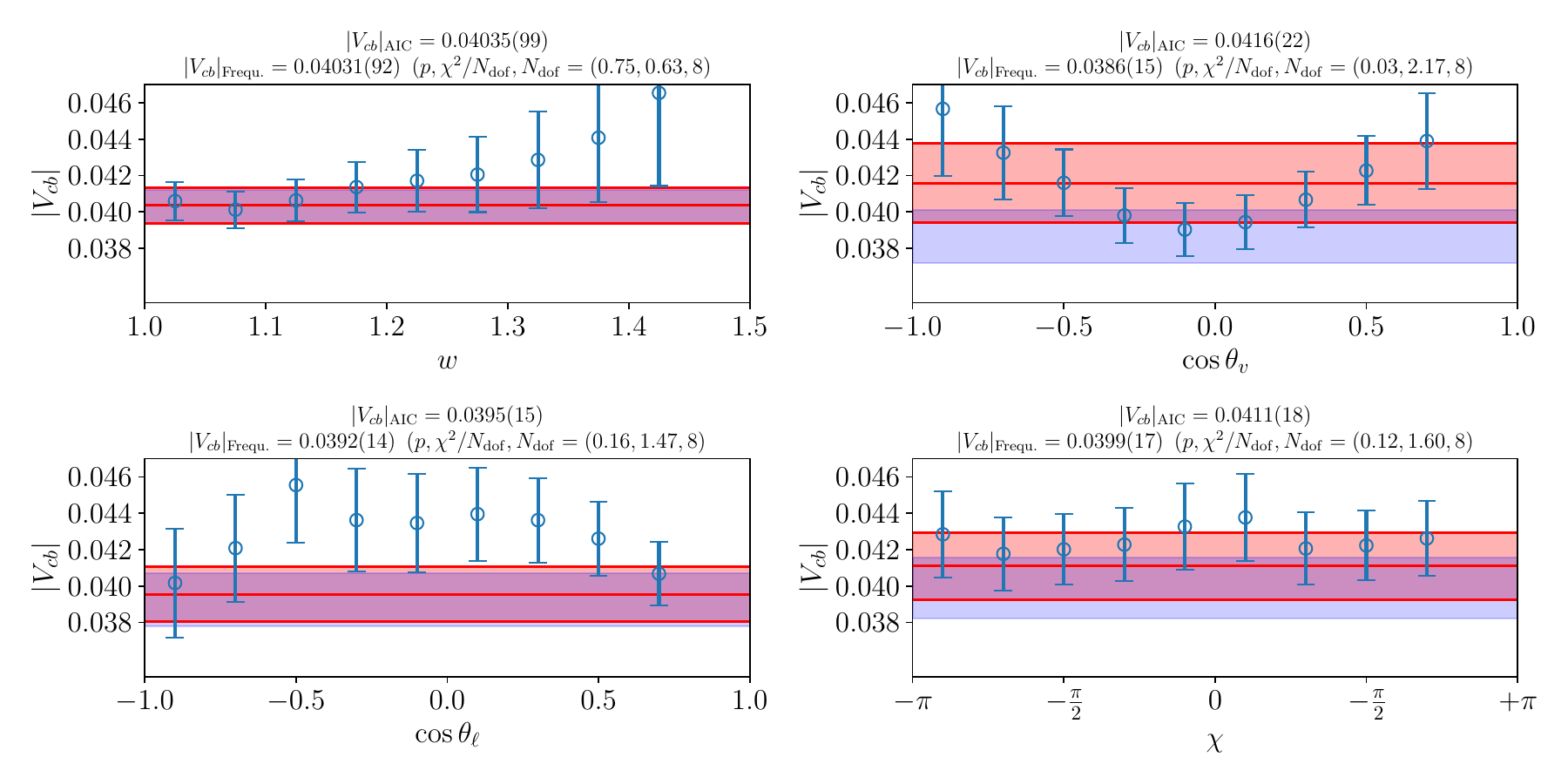}\\
        JLQCD 23\\
        \includegraphics[width=12cm]{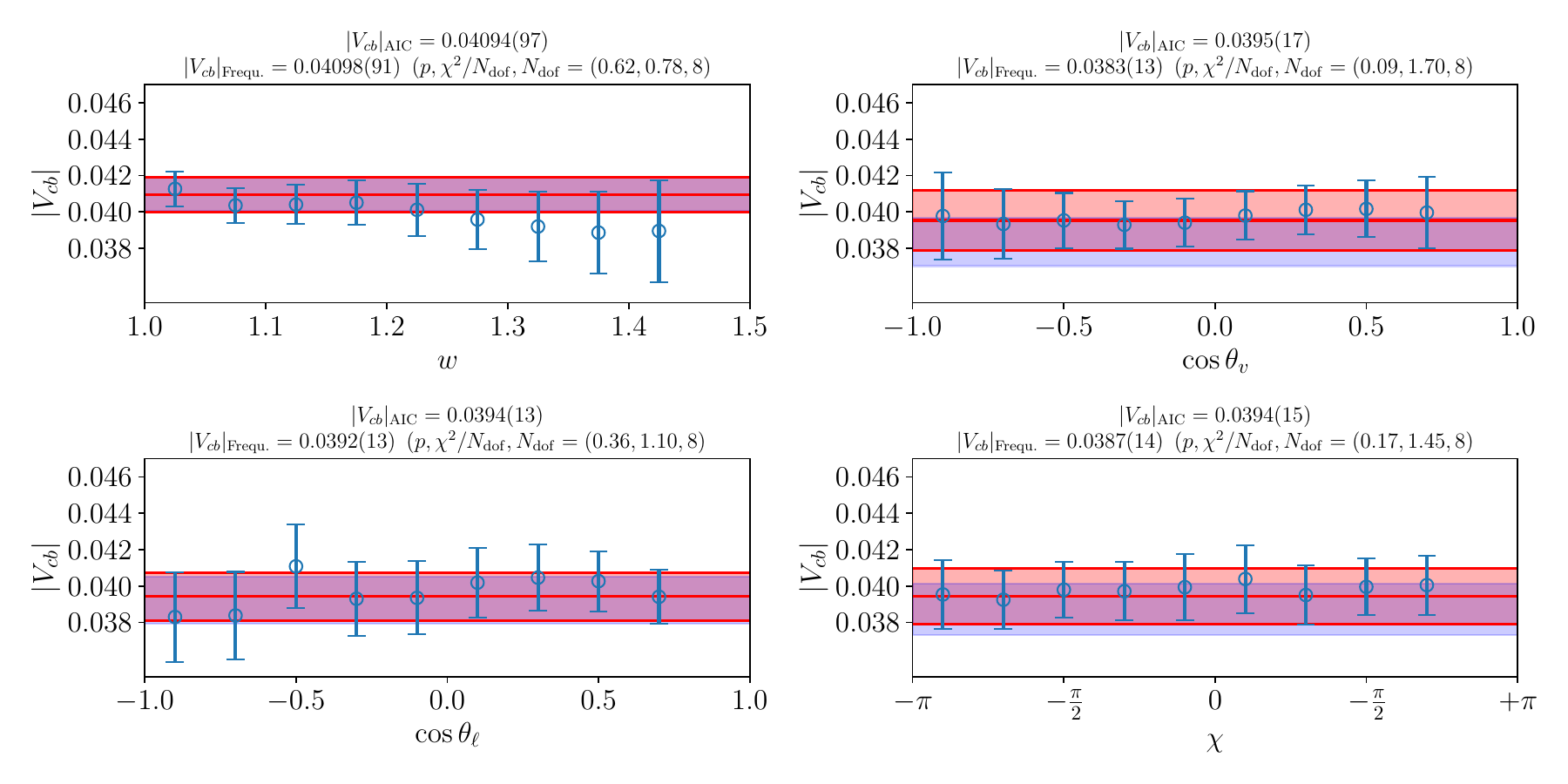}
    \end{center}
    \caption{$|V_{cb}|_{X,i}$ (cf. Eq.~\ref{eq:Vcb_bin})) for  $X=w,\,\cos\theta_v,\,\cos\theta_\ell$ and $\chi$ for HFLAV 24 data and combined BGL $(K_f,K_\Fone,K_\Ftwo,g)=(4,4,4,4)$ Bayesian fit to lattice results
    from FNAL/MILC 21, HPQCD 23 and JLQCD 23, respectively. Blue data points are results on a given experimental bin, horizontal red band in each case is the AIC-average (cf. Eqs.~(\ref{eq:AIC weight})-(\ref{eq:AIC error})), and
    the blue band in each case corresponds to the naive frequentist fit to all shown data points. See plot titles for central values and fit quality in the case of the frequentist fit.}\label{fig:Vcb bin-by-bin-by-X}
\end{figure}

The bin-by-bin results for $\Vcb$ following Eq.~(\ref{eq:Vcb_bin}) are shown in Fig.~\ref{fig:Vcb bin-by-bin-by-X} based on HFLAV~24 and BGL fits to individual lattice data sets. Somewhat surprisingly, but in agreement with the findings of~\cite{Martinelli:2021onb,Martinelli:2023fwm} based on the dispersive-matrix method, the results for $\Vcb_{X,i}$ from   FNAL/MILC~21 and HPQCD~23 vary significantly from bin to bin. Each plot also shows the result of a naive frequentist as well as the AIC fit to all shown data points as blue and red bands, respectively. The $p$ values shown in the title for the correlated constant fit to all bins indicate that not all fits are of acceptable quality. In particular, the results based on the angular differential decay rates are of comparatively low quality. Furthermore, the frequentist fits, for which we often find low $p$ values,
in some cases appear to suffer from the d'Agostini bias \cite{DAgostini:1993arp}, shifting the central value  for $\Vcb$ to values that are lower than the individual data points. These problems, which were also previously found in~\cite{Gambino:2019sif,Ferlewicz:2020lxm,Martinelli:2023fwm},   motivated us to use the AIC to formulate an alternative analysis strategy. The red bands describe the data better, and where still present (in particular FNAL/MILC~21 and HPQCD~23), the bias is much reduced. Note that in contrast to~\cite{Martinelli:2021onb,Martinelli:2023fwm}, only fits of acceptable quality enter the further analysis, and no PDG inflation~\cite{ParticleDataGroup:2022pth}  is required at intermediate steps.
\begin{figure}
    \begin{center}        \includegraphics[width=13cm]{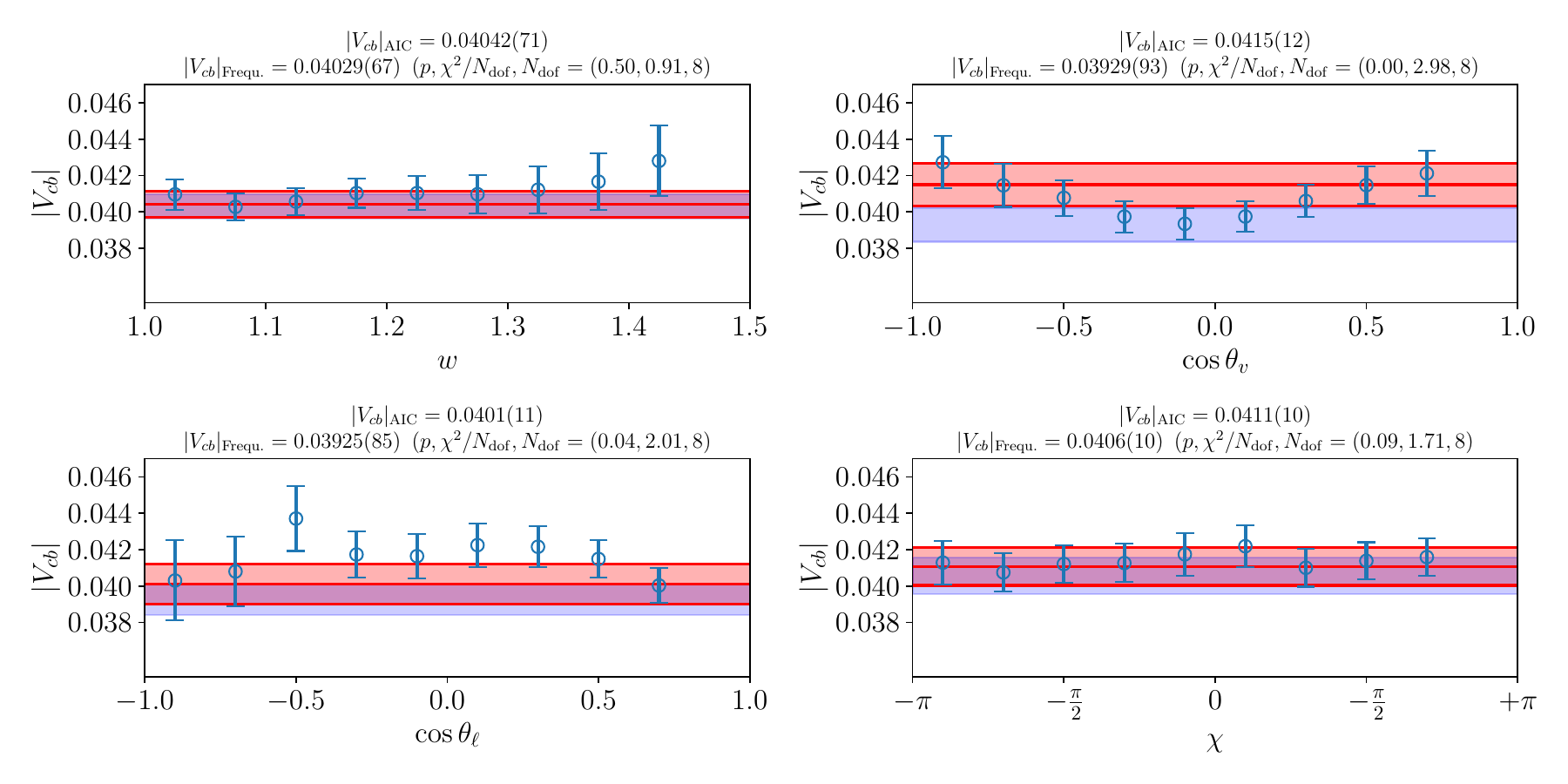}
    \end{center}
    \caption{$|V_{cb}|_{X,i}$ (cf. Eq.~\ref{eq:Vcb_bin})) for $X=w,\,\cos\theta_v,\,\cos\theta_\ell$ and $\chi$ for HFLAV 24 data and combined BGL $(K_f,K_\Fone,K_\Ftwo,g)=(4,4,4,4)$ Bayesian fit to lattice results
    from FNAL/MILC 21, HPQCD 23 and JLQCD 23. Blue data points are results on a given experimental bin and horizontal red band is the AIC-average (cf. Eqs.~(\ref{eq:AIC weight})-(\ref{eq:AIC error})) for $|V_{cb}|$ from a simultaneous analysis of all four channels. See plot titles for central values and fit quality in the case of the frequentist fit.}\label{fig:Vcb bin-by-bin}
\end{figure}
\begin{figure}
    \begin{center}
       \scalebox{0.6}{\input{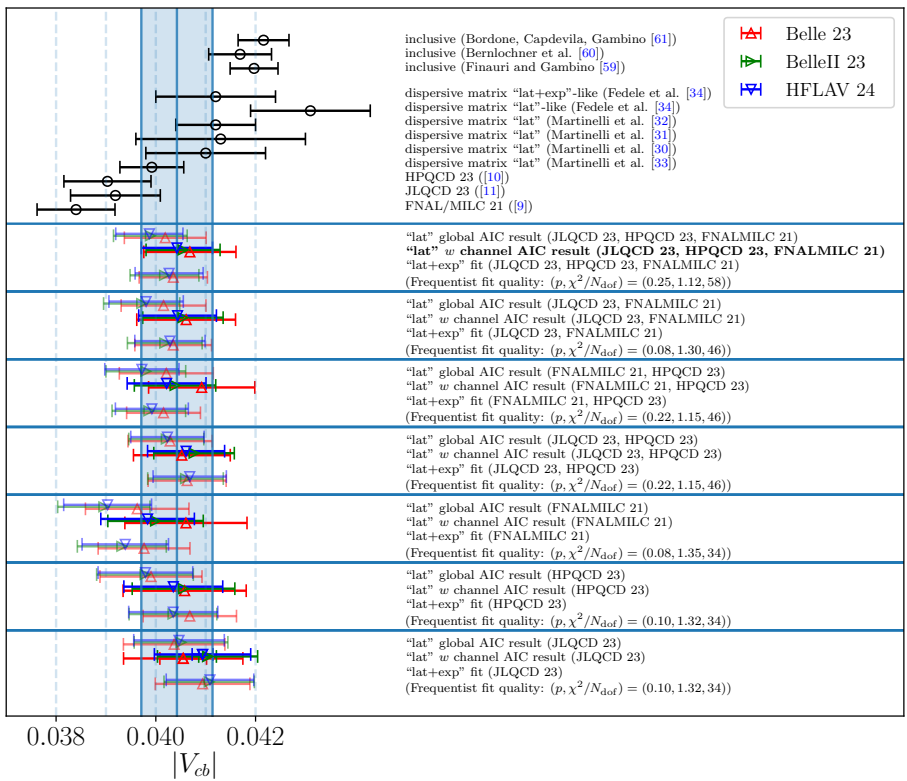}}
    \end{center}
    \caption{$\Vcb$ results obtained from the analyses in this paper and comparison with the literature (top). The results from the AIC analysis based on  the $w$ differential decay rate are shown in stronger colours. The blue band corresponds to our nominal result as in Eq.~(\ref{eq:res_Vcb}), where we employ the results from the $w$ distribution for the combined fit to all LQCD data sets and the HFLAV 24 experimental average. The dashed vertical lines are shown only to guide the
eye.}\label{fig:Vcb scatter}
\end{figure}
The extraction of $\Vcb$ is most consistent in all channels when based on the JLQCD 23 data set.
We draw the following conclusions:
\begin{itemize}
    \item There are issues with interpreting the bins based on the angular variables $\cos\theta_v\,,\cos\theta_\ell$ and $\chi$ based on the lattice input by FNAL/MILC 21 and HPQCD 23 -- this is not visibly the case for JLQCD 23.
    \item The results based on the $w$ bins look more consistent with very good $p$ values, and the results of the naive frequentist fit and AIC are in agreement. 
\end{itemize}
Despite the above observations, we find that all individual fit results based on the $w$ distribution, or also including the three angular distributions, are compatible within two standard deviations, as illustrated in Fig.~\ref{fig:Vcb scatter}. Nevertheless, the poor fit quality that we find in the bin-by-bin results for each of the one-dimensional angular distributions motivates us to discard them, and focus only on the $w$ distribution to extract $\Vcb$. We use the combined fit to all three lattice data sets together with the HFLAV 24 combination of experimental data in the AIC framework, and find:
\begin{equation}
\parbox[l]{3.4cm}{\centering\textrm{``lat'' analysis\\ $w$ bins}}\quad\,\Vcb=0.04025(71)\quad\,
\begin{array}{ll}
     \textrm{FNAL/MILC 21~\cite{FermilabLattice:2021cdg},}         &\textrm{Belle 23~\cite{Belle:2023bwv,hepdata.137767},}\\
     \textrm{HPQCD 23~\cite{Harrison:2023dzh},}     &\textrm{Belle II 23~\cite{Belle-II:2023okj,hepdata.145129},}\\
     \textrm{JLQCD 23~\cite{Aoki:2023qpa},}&\textrm{HFLAV 24~\cite{HFLAV:2024}\,.}\\
\end{array}
\label{eq:res_Vcb}
\end{equation}
The corresponding frequentist fit leads to $\Vcb=0.04012(66)$ with $(p,\chi^2/N_{\rm dof},N_{\rm dof})=(0.5,0.91,8)$.
\subsubsection{$\Vcb$ from the ``lat+exp'' fit}
Results for the combined fit to experimental and lattice data discussed in Sec.~\ref{sec:exp+lat fit} are illustrated in Figs.~\ref{fig:BGL fit example} and \ref{fig:dGdX}, respectively, and the results for $\Vcb$ are shown in Fig.~\ref{fig:Vcb scatter}. The latter plot also provides the quality of fit in each case. 
Apart from the fit with FNAL/MILC 21, there is good compatibility of all results within one standard deviation. Repeating the note of caution that the analysis leading to this result imposes SM constraints on the shape of the experimental differential decay rate, our
overall result for $\Vcb$ therefore is the one based on HFLAV 24 combined with FNAL/MILC 21, HPQCD 23 and JLQCD 23:
\begin{equation}\label{eq:Vcb global fit}
\parbox[l]{3.2cm}{\centering\textrm{``lat+exp'' analysis\\all bins}}\quad\,\Vcb=0.04037(74)\quad\,
\begin{array}{ll}
     \textrm{FNAL/MILC 21~\cite{FermilabLattice:2021cdg},}         &\textrm{Belle 23~\cite{Belle:2023bwv,hepdata.137767},}\\
     \textrm{HPQCD 23~\cite{Harrison:2023dzh},}     &\textrm{Belle II 23~\cite{Belle-II:2023okj,hepdata.145129},}\\
     \textrm{JLQCD 23~\cite{Aoki:2023qpa},}&\textrm{HFLAV 24~\cite{HFLAV:2024}\,.}\\
\end{array}
\end{equation}
For the corresponding frequentist fit we find $(p,\chi^2/N_{\rm dof},N_{\rm dof})=(0.13,1.21,58)$.
At this level of precision we see no deviations with respect to the results of the previous sections. And hence, as far as $\Vcb$ is concerned, the SM assumptions entering the BGL fit are compatible with the shape of the differential decay rates.

\subsubsection{Discussion}
The central result for $\Vcb$ in this paper, Eq.~(\ref{eq:res_Vcb}), is shown as vertical blue band in Fig.~\ref{fig:Vcb scatter}. At the top of this scatter plot we also show results from other analyses of both exclusive and inlcusive decays. The dispersive-matrix analyses in~\cite{Martinelli:2023fwm,Martinelli:2021onb,Martinelli:2021myh,Martinelli:2022xir,Fedele:2023ewe} are close in spirit to ours, in that both apply the unitarity constraint within the dispersive-matrix approach. The work in~\cite{Martinelli:2023fwm}, which on top of Belle 23 and Belle II 23 also includes the earlier Belle data set~\cite{Belle:2018ezy}, and is the most recent in a series of papers~\cite{Martinelli:2021onb,Martinelli:2021myh,Martinelli:2022xir}, leads to a result that is fully compatible with ours. Their analysis within the ``lat'' approach includes results from all three lattice collaborations, and for differential decay rates for all channels $w$, $\cos\theta_\ell$, $\cos\theta_v$ and $\chi$. In order to mitigate the inconsistencies in the angular decay rate also found in our work (see Sec.~\ref{sec:Vub lat}), they employ a PDG scaling factor. Their final result is $\Vcb=0.03992(64)$. The  work in~\cite{Fedele:2023ewe} is somewhat closer  in spirit, but not the same, as the ``lat+exp'' analysis. They use the results for form factors from a dispersive-matrix ``lat'' analysis of FNAL/MILC 21 data as priors in a fit to  experimental-decay-rate data from the earlier  Belle~\cite{Belle:2017rcc,Belle:2018ezy} data set. The value for $\Vcb$ is then obtained from the integrated decay rate, once based on their ``lat'' fit, and once with the integrated decay rate from the ``lat+exp''-type fit. The results from this analysis have larger statistical errors and central values lie  higher than the other dispersive-matrix results by~\cite{Martinelli:2023fwm}. 
While based on our results and the ones of~\cite{Martinelli:2023fwm}, a small tension with the inclusive determinations of~\cite{Finauri:2023kte,Bernlochner:2022ucr,Bordone:2021oof} persists, the analysis of~\cite{Fedele:2023ewe} concludes that inclusive and exclusive determinations are compatible.
In order to better understand this scatter in results, the conclusions from the dispersive-matrix approach, and what this means for the comparison to our analysis, it would be good to have a repeat of the analysis in~\cite{Fedele:2023ewe}, but including all three lattice data sets, as well as the newest experimental data. Moreover, and if practicable, it would be interesting to compare to a full dispersive-matrix analysis that simultaneously uses lattice and experimental data as input, i.e., along the lines of ``lat+exp'', but without the use of the ``lat'' fit as prior.

Fig.~\ref{fig:Vcb scatter} also shows the results that were published by the three lattice collaborations. For FNAL/MILC 21 the result is based on experimental input by Belle~\cite{Belle:2018ezy} and BaBar~\cite{BaBar:2019vpl}, for HPQCD 23 on Belle~\cite{Belle:2018ezy}, and for JLQCD 23 also on Belle~\cite{Belle:2018ezy}. These results are lower than ours by up to slightly above one standard deviation. A more comprehensive comparison, which is beyond the scope of this paper, would also include results from, e.g., the averaging groups FLAG~\cite{FlavourLatticeAveragingGroupFLAG:2021npn}, PDG~\cite{ParticleDataGroup:2022pth}, UTfit~\cite{UTfit:2022hsi}, CKMFitter~\cite{ValeSilva:2024jml}.

\subsection{Other observables}
Further to the forward-backward asymmetry $A^\ell_{\rm FB}$ and the longitudinal $D^\ast$ polarisation fraction $F_L^\ell$ defined in Eqs.~(\ref{eq:AFB}) and (\ref{eq:FL}), respectively, we also consider the LFU ratio 
\begin{equation}
R^{\ell_1/\ell_2}(D^\ast) = {\frac{\mathcal{N}_{m_{\ell_1}}}{\mathcal{N}_{m_{\ell_2}}}}\,,
\end{equation}
\begin{figure}
    \centering
   \scalebox{0.52}{\input{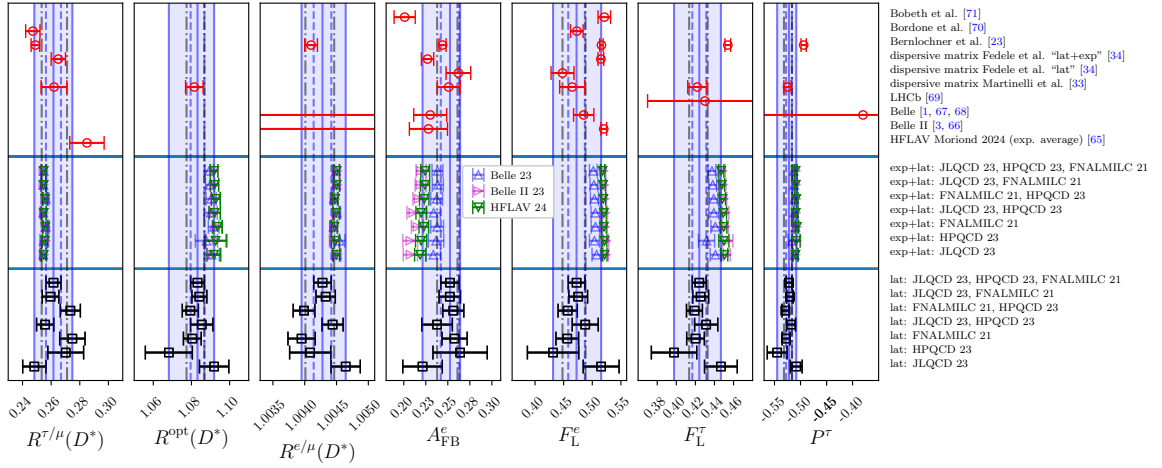}}
    \caption{{Comparison of results  based on BGL fits to lattice data (squares), simultaneous BGL fit to experiment (see legend) and lattice data (triangles), fits to only experimental data (circles) by Belle~\cite{Belle:2019ewo,Belle:2023bwv,hepdata.137767,Belle:2023xgj}, Belle II~\cite{Belle-II:2023okj,hepdata.145129,Belle-II:2023svm}, LHCb~\cite{LHCb:2023ssl} and the HFLAV-Moriond~24-average~\cite{HFLAVRDstar:2024,BaBar:2012obs,BaBar:2013mob,Belle:2015qfa,Belle:2016dyj,Belle:2017ilt,Belle:2019rba,LHCb:2023zxo,LHCb:2023uiv,Belle-II:2024ami,LHCbc:2024}, where available or visible within the shown range along the horizontal axis, and other work we comment on in the text (also circles). All shown BGL fits are for $(K_f,K_{\Fone},K_\Ftwo,K_g$)=(4,4,4,4). The vertical blue band indicates our central results presented in Eq.~(\ref{eq:Ratios central lat}), the vertical dashed line is the corresponding statistical error, and the dash-dotted grey lines indicate the range for the results that we would obtain using PDG error inflation (see text after Eq.~(\ref{eq:Ratios central lat})).}}
    \label{fig:pheno scatter}
\end{figure}
the optimised ratio \cite{Isidori:2020eyd}
\begin{equation}
    {R^\mathrm{opt} = \frac
    {I_{m_\tau}[\omega_\tau\{H_0^2 + H_-^2 + H_+^2+\rho_\tau(q^2)H_S^2\}]}
    {I_{m_\tau}[\omega_\tau \{H_0^2 + H_-^2 + H_+^2\}]} \,,}
\end{equation}
where $\omega_\ell(q^2)=(1+m_\ell^2/(2q^2))$ and $\rho_\tau(q^2)=3m_\ell^2/(m_\ell^2+2q^2)$, and the $\tau$ polarisation~\cite{Fajfer:2012vx}
\begin{equation}
   { P^\tau=I_{m_\tau}\left[\left(1+\frac{m_\tau^2}{2q^2}\right)\left(H_0^2 + H_-^2 + H_+^2\right)-\frac 32\frac{ m_\tau^2}{q^2} H_S^2\right]/\mathcal{N}_{m_\tau}}
\end{equation}

Our predictions for the integrated observables are illustrated in Fig.~\ref{fig:pheno scatter} and numerical values are given in  Tab.~\ref{tab:pheno observables}.
\begin{table}[]
    \begin{center}
    \resizebox{1\textwidth}{!}{  
    \begin{tabular}{lllllllllllllllllllllllllllllllllllllllllllllllllll}
\hline\hline
lat & \multicolumn{1}{c}{$R^{\tau/\mu}(D^\ast)$}&		\multicolumn{1}{c}{$R^{\rm opt}(D^\ast)$}&		\multicolumn{1}{c}{$R^{e/\mu}(D^\ast)$}&		\multicolumn{1}{c}{$A^e_{\rm FB}$}&		\multicolumn{1}{c}{$F^e_{\rm L}$}&		\multicolumn{1}{c}{$F^\tau_{\rm L}$}&		\multicolumn{1}{c}{$P^\tau$} \\
\hline
JLQCD 23&0.2482(81)&1.0919(77)&1.00464(23)&0.221(22)&0.515(31)&0.447(17)&-0.508(11)\\
HPQCD 23&0.270(13)&1.068(12)&1.00409(32)&0.264(31)&0.432(45)&0.398(24)&-0.545(19)\\
FNAL/MILC 21&0.2748(89)&1.0805(47)&1.00395(21)&0.258(14)&0.456(20)&0.4202(93)&-0.5277(74)\\
JLQCD 23 HPQCD 23&0.2558(60)&1.0854(59)&1.00444(17)&0.238(17)&0.488(23)&0.431(12)&-0.5183(87)\\
FNAL/MILC 21 HPQCD 23&0.2734(70)&1.0794(42)&1.00399(17)&0.256(12)&0.457(17)&0.4191(83)&-0.5290(66)\\
JLQCD 23 FNAL/MILC 21&0.2596(58)&1.0841(39)&1.00433(15)&0.252(12)&0.475(16)&0.4255(84)&-0.5204(60)\\
JLQCD 23 HPQCD 23 FNAL/MILC 21&0.2616(52)&1.0832(36)&1.00428(14)&0.252(10)&0.473(15)&0.4241(73)&-0.5221(56)\\
\hline
lat+exp & \multicolumn{1}{c}{$R^{\tau/\mu}(D^\ast)$}&		\multicolumn{1}{c}{$R^{\rm opt}(D^\ast)$}&		\multicolumn{1}{c}{$R^{e/\mu}(D^\ast)$}&		\multicolumn{1}{c}{$A^e_{\rm FB}$}&		\multicolumn{1}{c}{$F^e_{\rm L}$}&		\multicolumn{1}{c}{$F^\tau_{\rm L}$}&		\multicolumn{1}{c}{$P^\tau$} \\
\hline
JLQCD 23&0.2548(17)&1.0918(36)&1.004497(52)&0.2187(64)&0.5215(42)&0.4505(35)&-0.5096(49)\\
HPQCD 23&0.2556(20)&1.0927(55)&1.004483(67)&0.2197(64)&0.5213(42)&0.4499(53)&-0.5085(76)\\
FNAL/MILC 21&0.2560(16)&1.0937(25)&1.004470(45)&0.2227(55)&0.5203(40)&0.4497(33)&-0.5070(34)\\
JLQCD 23 HPQCD 23&0.2549(16)&1.0922(30)&1.004495(48)&0.2197(59)&0.5203(40)&0.4493(34)&-0.5090(41)\\
FNAL/MILC 21 HPQCD 23&0.2558(16)&1.0928(23)&1.004479(44)&0.2232(54)&0.5193(39)&0.4484(32)&-0.5082(32)\\
JLQCD 23 FNAL/MILC 21&0.2548(15)&1.0921(22)&1.004502(43)&0.2241(53)&0.5188(39)&0.4476(29)&-0.5091(30)\\
JLQCD 23 HPQCD 23 FNAL/MILC 21&0.2548(15)&1.0919(20)&1.004503(42)&0.2243(50)&0.5179(38)&0.4470(29)&-0.5094(28)\\
\hline\hline\\
\end{tabular}
}
    \caption{Summary of results. The top panel of the table is based on BGL fits to only lattice data, while the bottom panel is based on combined BGL fits to lattice and experimental data (HFLAV 24). A summary of these results is also provided in Fig.~\ref{fig:pheno scatter}.}
    \label{tab:pheno observables}
    \end{center}
\end{table}
\subsubsection{Other observables from the ``lat'' fit}
The bottom panel shows predictions from BGL fits ``lat'' to different combinations of lattice data sets, i.e., independent of experiment. For $R^{\tau/\mu}(D^\ast)$ we observe that FNAL/MILC~21 and HPQCD 23 prefer larger values than JLQCD 23. This leads to a tension of two to three standard deviations amongst the data points. Nevertheless, the results based on combined fits are, as discussed in Sec.~\ref{sec:lat fit}, of acceptable quality. We observe a similar pattern (or its inverse) for the other observables shown in Fig.~\ref{fig:pheno scatter}. The observed scatter can likely be traced back to the different shapes of the parameterisations of  individual lattice form factors illustrated in terms of grey bands in Fig.~\ref{fig:BGL fit example}. For instance, the observables $F_L^\ell$ depends on the helicity form factor $H_0^2(w)=\Fone(w)/q^2$, and $\Fone(w)$ from JLQCD 23 has a milder slope with $w$ compared to FNAL/MILC 21 and HPQCD 23. 

Given this scatter, identifying a best-fit result is not straight-forward. Being cautious, we use the combined fit to all three lattice results and attach a systematic error such that the total error reduces the tension with the central values of individual fit results to the level of one standard deviation:
\begin{align}
R^{\tau/\mu}(D^\ast) &= 0.262(5)(12), &
R^{e/\mu}(D^\ast) &= 1.00428(14)(^{+34}_{-30}),\nonumber\\
R^{\rm opt}(D^\ast) &= 1.083(4)(^{+8}_{-15}),&
A^e_{\rm FB} &= 0.252(10)(^{+5}_{-29}) ,\nonumber\\
F^e_{\rm L} &= 0.473(15)(^{+40}_{-38}), &
F^\tau_{\rm L} &= 0.424(7)(^{+21}_{-25}), \nonumber\\
P^\tau &= -0.522(6)(^{+13}_{-22})\,,\label{eq:Ratios central lat}
\end{align}
where the first error is statistical and the second error is systematic as described. Had we instead obtained the central values from a constant fit to the three individual lattice results together with PDG inflation~\cite{ParticleDataGroup:2022pth}, we would obtain
$R^{\tau/\mu}(D^\ast)=0.262(9)$,
$R^{e/\mu}(D^\ast)=1.00424(23)$, 
$R^{\rm opt}(D^\ast)=1.082(5)$, 
$A^e_{\rm FB} =0.249(12)$, 
$F^e_{\rm L}=0.468(20)$, 
$F^\tau_{\rm L}=0.423(10)$, 
$P^\tau=-0.524(8)$. These results are compatible with our central results, but with smaller errors.

\subsubsection{Other observables from the  ``lat+exp'' fit}
Moving on to the combined BGL ``lat+exp'' fit over lattice and experimental results and referring again to Fig.~\ref{fig:pheno scatter}, the agreement for $R^{\tau/\mu}(D^\ast)$ under variation of the lattice as well as experimental input is striking. At the same time, the result is in agreement with the lattice-only result in Eq.~(\ref{eq:Ratios central lat}). It appears that, for this particular quantity, the shape information provided by the experimental input smoothens out any tension observed in the lattice-only fits.  {Note that the predictions for $R^{\tau/\mu}(D^\ast)$ and other observables depending on the $\tau$-lepton mass are based on the SM expressions, but with BGL coefficients from fits to the experimental data assuming $m_\ell=0$.} A similar agreement of results under variations of the lattice input is also observed for other observables shown in the plot.
However, a significant tension between {the predictions based on} Belle 23 and Belle II 23 becomes apparent for $A^e_{\rm FB}$$, F^e_{\rm L}$ and $F^\tau_L$. It could be that the larger error bars in Belle~23's differential decay rate in these cases allows the lattice data to \emph{pull} the central value towards the results for the ``lat'' fit. However, comparing to the top panel of Fig.~\ref{fig:pheno scatter}, one finds that the respective Belle 
\cite{Belle:2023xgj} and Belle-II 
\cite{Belle-II:2023svm} dedicated analyses for $F_{\rm L}^e$  show a similar tension between the two experiments. For $A_{\rm FB}^e$ no such tension is currently seen in the experimental results.

\subsubsection{Discussion}
Within the comparatively large combined statistical and systematic uncertainties of our central results in Eq.~(\ref{eq:Ratios central lat}), we find compatibility of the ``lat'' and ``lat+exp'' analyses. The somewhat surprising scatter of results in the ``lat'' analysis of individual lattice results motivates this error. The scatter of lattice result could indeed be a statistical fluctuation only, but the shifts in BGL coefficients that we found indicates, that there might be some issue with the $w$ dependence of the lattice form factors. Future simulations will hopefully shed light on this tension, and then allow us to reduce this systematic error.

Regarding the ``exp+lat'' analysis, we find consistency between fits to different lattice input and also to different experimental input, except for $A_{\rm FB}^e$, $F_{\rm L}^e$ and $F_{\rm L}^\tau$, where results based on different experimental input are at significant tension. This is intriguing and requires further scrutiny. The fit to the combined data set HFLAV 24 ends up lying either in between, or closer to the Belle II 23 result, which does also have smaller errors for the differential decay rates.

In the top panel of Fig.~\ref{fig:pheno scatter} we also show results from experiment, dispersive-matrix and other determinations. These results are compatible at the 1$\sigma$ level (in some cases slightly more) with ours. We note that the ``lat'' results of~\cite{Martinelli:2023fwm} agree almost exactly with ours, when we use PDG inflation (see results after Eq.~(\ref{eq:Ratios central lat})). This is comforting and a valuable consistency check. Similarly, where available, we agree with the results for $A_{\rm FB}^e$ and $F_{\rm L}^e$ of the ``lat+exp''-type study of~\cite{Fedele:2023ewe}, while there is a bit of a tension in the case of $R^{\tau/\mu}(D^\ast)$.
\section{Summary and Conclusions}
In this work, we study the recent determination of $\BtoDstarellnu$ hadronic form factors from FNAL/MILC 21~\cite{FermilabLattice:2021cdg}, HPQCD 23~\cite{Harrison:2023dzh} and JLQCD 23~\cite{Aoki:2023qpa} in light of two new data sets from Belle~\cite{Belle:2023bwv,hepdata.137767} and Belle II~\cite{Belle-II:2023okj,hepdata.145129} and, for the first time, their combination by HFLAV 24~\cite{HFLAV:2024}. We study at length the compatibility of the three lattice data sets, fitting a BGL parametrisation for the hadronic form factors using the Bayesian approach as in \cite{Flynn:2023nhi,Flynn:2023qmi}, as well as using a frequentist fit. All three LQCD data sets yield a good (in the frequentist sense) fit to a BGL parametrisation for the hadronic form factors, but we notice some differences in the results for the fit parameters especially between the JLQCD 23 on the one hand, and FNAL/MILC 21 or HPQCD 23 on the other. These differences are of the order of a few sigmas, with the larger one being $2.5\sigma$ for $a_{\Ftwo,1}$. Nevertheless, we also find that the combined fit between LQCD  data sets and experimental data yields fits of acceptable quality.

We use these studies for phenomenological purposes and first extract $\Vcb$. We find that a bin-by-bin analysis for $\Vcb$, where one first fits a BGL ansatz to the lattice data and then combines the results with experimental bins, shows tensions in the angular distributions when based on the FNAL/MILC 21 and HPQCD 23 data sets. No such tensions are observed in the case of JLQCD 23. We find poor quality of fits when combining the bin-by-bin values for $\Vcb$, and the results appear to suffer {from the} d'Agostini bias \cite{DAgostini:1993arp}. As a mitigation measure we employ a weighted average over all possible sub-sets of fits, based on the Akaike information criterion~\cite{1100705,gamage2016adjusted}. We restrict the further analysis to the $w$ channel, where no tensions are found, leading to our final result
\begin{equation}
    \Vcb = 0.04025(71)\,.
\end{equation}
This result is based on the combined Bayesian BGL fit to all three lattice data sets~\cite{FermilabLattice:2021cdg, Harrison:2023dzh,Aoki:2023qpa}, and on the combination of Belle 23~\cite{Belle:2023bwv,hepdata.137767} and Belle II 23~\cite{Belle-II:2023okj,hepdata.145129} data by HFLAV 24~\cite{HFLAV:2024}. It is compatible with our simultaneous fit to lattice and experimental data (``lat+exp''), as seen from Eq.~(\ref{eq:Vcb global fit}). Together with the absence of any tension in this fit, which imposes the SM assumptions on the form-factor shape inherent in the BGL parameterisation onto the experimental data, this indicates no NP contributions at the current level of precision. Our result is also compatible with the combined fit for the inclusive determination of $\Vcb$ \cite{Finauri:2023kte} (see also \cite{Bordone:2021oof,Bernlochner:2022ucr}) at the $2\sigma$ level. 
%In comparison with the recent dispersive-matrix analysis~\cite{Martinelli:2023fwm}, as illustrated in Fig.~\ref{fig:Vcb scatter}, our result has slightly larger uncertainties, which, however, we believe to better represent our current knowledge of the theoretical predictions for $\BtoDstarellnu$ hadronic form factors. \\

We also predict other phenomenologically relevant observables, and compare them to the experimental measurements and previous literature. The summary of our results can be found in  Fig.~\ref{fig:pheno scatter}, which highlights some inconsistencies. First, the fits to only lattice data (``lat'') show a spread in the predictions for all the observables, for instance for $R^{\tau/\mu}(D^*)$, depending on which lattice-data set was used as input (as also reported in~\cite{Martinelli:2023fwm}). Nevertheless, we choose to fix the central value for our prediction to the ones from the combined fit to all three LQCD data sets, which exhibits acceptable frequentist quality of fit. We then, however, add a systematic error that accounts for the aforementioned spread. As a result, our nominal predictions suffer from larger uncertainties than other analyses that are based only on theory input and not experimental data (i.e., ``lat'').
Second, in fits to both lattice and experimental data (``lat+exp''), we find that the forward-backward asymmetry $A^e_{\rm FB}$ and the $D^\ast$ polarisation fractions $F_{\rm L}^{e}$ and $F_{\rm L}^\tau$ yield incompatible results depending on whether the  Belle~2023 or Belle~II~2023 data sets are used, regardless of the LQCD data employed. This points to a tension between the current Belle 2023 and Belle II 2023 data sets that will hopefully be understood with future experimental data and analyses. 
These results are complementary to the ``lat'' analysis, and provide additional information that might eventually help to track down NP signals. The complementary use of a frequentist analysis in this context is valuable, since the $p$ value provides an important indicator of tensions between data and experiment.

Our analyses reveal some tensions amongst theory expectations and experimental data. Nonetheless,  with the current sensitivity, it is not possible to draw any firm conclusion about their origin. In light of this, and the prospects of new experimental data as well as foreseen improvements in the theory computations, the study of $\BtoDstarellnu$ remains essential and provides exciting perspectives. The analysis strategies proposed and demonstrated in this work will help further constraining the SM with the study of not only $\BtoDstarellnu$ decays, but also applied to other exclusive semileptonic decay channels.
To end, we highlight the ease with which the experimental and lattice data could be combined within the ``lat+exp'' analysis within the Bayesian-inference framework of~\cite{Flynn:2023qmi}.

\acknowledgments We particularly thank Florian Bernlochner and Markus Prim for discussions and for providing early-access to the combination of Belle~23 and Belle~II~23 data, labelled HFLAV~24 in the text. We also thank our collaborators Tobias Tsang and Jonathan Flynn for discussions around the Bayesian-inference framework and for very valuable comments on the manuscript. We have made use the \verb|NumPy|~\cite{harris2020array}, \verb|SciPy|~\cite{2020SciPy-NMeth} and \verb|Matplotlib|~\cite{Hunter:2007}, \verb|PyMultiNest|~\cite{Buchner:2014nha} and \verb|BFF|~\cite{andreasjuettner_2023_7799451} Python libraries.

\appendix
\section{Further details on the BGL implementation}
\label{app:A}
The matrix $Z$ has the block-diagonal entries
\begin{equation}
    (Z_{XX})_{ij}=\frac{1}{\phi_X(z_i)B_X(z_i)}(z_i)^{j}\,,
\end{equation}
for \(X=f,\Fone, \Ftwo, g\). The  index \(i\)
runs over the available discrete \(z\) values, while the index \(j\)
contracts with the elements of the parameter vector ${\bf }$.
The kinematic constraints constraints Eqs. (\(\ref{eq:constraint1}\)) and (\(\ref{eq:constraint2}\)) are implemented in terms of the following off-diagonal
blocks:
\begin{align}
(Z_{f,\mathcal{F}_1})_{i0}&=\frac{1}{M_B(1-r)}\frac{\phi_f(z(q^2_{\rm max}))}{\phi_{\mathcal{F}_1}(z(q^2_{\rm max}))}\frac{1}{\phi_f(z_i)B_f(z_i)}\,,\\
(Z_{\mathcal{F}_2,\mathcal{F}_1})_{ij}&=
\frac{1+r}{M_B^2 r(1-r)(w_{\rm max}+1)}\frac{\phi_{\mathcal{F}_2(z_{\rm max})}B_{\mathcal{F}_2}(z_{\rm max})}{\phi_{\mathcal{F}_1(z_{\rm max})}B_{\mathcal{F}_1}(z_{\rm max})}\frac{1}{\phi_{\mathcal{F}_2}(z_i)B_{\mathcal{F}_2}(z_i)}(z_i)^j\,.
\end{align}

\section{Lattice data sets}\label{app:Lattice data sets}
To date, results from three different collaborations are available, each using different discretisations of QCD and independent sets of gauge configurations. We therefore consider results from different collaborations as statistically independent. We briefly discuss basic properties and comment on data curation.
\subsection{Synthetic lattice data from JLQCD~\cite{Aoki:2023qpa}}
These results are based on $N_f=2+1$ flavours M\"obius Domain-Wall-Fermions~\cite{Brower:2012vk}, tree-level improved Symanzik gauge action~\cite{Weisz:1982zw,Weisz:1983bn,Luscher:1984xn} with lattice spacings in the range 0.08 -- 0.04fm and pion masses above 230MeV. Synthetic data for form factors $f$, $\Fone$, $\Ftwo$ and $g$ and their combined statistical and systematic covariance matrix at reference $w$ values 1.025, 1.060 and 1.100
are tabulated in Tab. IV of~\cite{Aoki:2023qpa}. The condition number of the corresponding correlation matrix is $1\cdot 10^4$.
\subsection{Synthetic lattice data from HPQCD~\cite{Harrison:2023dzh}}
The simulations are based on gauge ensembles of $N_f=2+1+1$ flavours of Highly-Improved-Staggered Quarks (HISQ)~\cite{Follana:2006rc} lattice spacings in the range 0.09 -- 0.045fm and pion masses above 135MeV. Synthetic data for form factors $h_V$, $h_{A_1}$, $h_{A_2}$ and $h_{A_3}$ and their covariance matrix at the kinematic points $w\approx 1.50, 1.38, 1.25, 1.13$  and $1.00$ are provided as supplementary material of~\cite{Harrison:2023dzh} in a binary format
that can be read by the python library \verb|gvar|~\cite{peter_lepage_2024_10797861}.
We use resampling to generate central values and covariances for the form factors $f$, $\Fone$, $\Ftwo$ and $g$ using the identities
\begin{align}
f       &= M_{B} \sqrt{r} (1+w) h_{A_1}\,,\\
\Fone      &= M_B^2 \sqrt{r}  (1+w)  \left((w-r)  h_{A_1} - (w-1) (r h_{A_2} + h_{A_3})\right)\,,\\
\Ftwo      &= \frac{1}{\sqrt{r}}  \left((1+w) h_{A_1} + (r w-1) h_{A_2} + (r-w) h_{A_3}\right)\,,\\
g       &= \frac{h_V}{M_B\sqrt{r}}\,.
\end{align}
One cause of concern is that both the data for $h_V$, $h_{A_1}$, $h_{A_2}$ and $h_{A_3}$ and $f$, $\Fone$, $\Ftwo$ and $g$ are highly correlated with condition numbers of the respective correlation matrices $10^{16}$ and $10^{15}$, respectively. We investigated various combination of pruning the data set. Considering now only the results for $f$, $\Fone$, $\Ftwo$ and $g$, removing the synthetic data for $w=1.5$ reduces the condition number to $10^{10}$, removing instead the results at $w=1.0$ does essentially not reduce the condition number. Removing both the results at $w=1.5$ and $w=1.0$, leaving us with three synthetic data points for each of the form factors $f$, $\Fone$, $\Ftwo$ and $g$, reduces
the condition number to a more acceptable $1\cdot 10^5$. We decided to base all analysis in this paper on 
the synthetic data points at  1.38, 1.25, 1.13 but note that other choices also lead to acceptably conditioned correlation matrices. 
\subsection{Synthetic lattice data from FNAL/MILC~\cite{FermilabLattice:2021cdg}}
Synthetic data for form factors $f$, $\Fone$, $\Ftwo$ and $g$ and their combined statistical and systematic covariance matrix at reference $w$ values 1.03, 1.10 and 1.17
are provided as supplementary material of~\cite{FermilabLattice:2021cdg} in \verb|gvar|~\cite{peter_lepage_2024_10797861} format. The simulations are based on $N_f=2+1$ flavours of asqtad-improved staggered sea quarks~\cite{MILC:2009mpl,Aubin:2004wf,Bernard:2001av} at five different lattice spacings in the range 0.15 -- 0.045fm, and pion masses above 180$
\,$MeV. The condition number of the corresponding correlation matrix is $3\cdot 10^4$.
\section{The experimental data sets}\label{app:Experimental data sets}
The Belle and Belle II collaborations provide data in terms of bins for the normalised differential decay rates $d\Gamma/d\alpha/\Gamma$ for $\alpha = \{w,\,\cos\theta_\ell,\,\cos\theta_v,\,\chi\}$. In the following, we comment on the differences between these data sets. 
\subsection{Experimental bins from Belle 23~\cite{Belle:2023bwv,hepdata.137767}}
This data set was obtained by the Belle collaboration based on their final integrated luminosity of 711~fb$^{-1}$ and using their improved hadronic tagging algorithm~\cite{Keck:2018lcd}. 
Results are therefore given in terms  \cite{Belle:2023bwv} of 40 bins, 10 for each kinematic variable~\cite{Belle:2023bwv}.
While the condition number of the $40\times 40$ combined statistical and systematic correlation matrix has a moderate condition number of $1\cdot 10^3$, we follow the same procedure as in Belle II 23~\cite{Belle-II:2023okj} (see below)
and discard for each $\alpha$ the last bin to take into account possible effects
from the normalisation of the differential decay rate. The condition number of the resulting $36\times 36$ correlation matrix is $30$.
\subsection{Experimental bins from Belle II 23~\cite{Belle-II:2023okj,hepdata.145129}}
The Belle II 23~\cite{Belle-II:2023okj} data set  is obtained from the analysis of experimental data at 189 fb$^{-1}$. Results are given in terms of 38 bins (10 in $w$,  8 in $\cos \theta_l$, 10 in $\cos\theta_v$ and 10 in $\chi$. However, since the number of events is the same for each kinematic distribution, the 38 bins are not independent. To avoid redundances, one bin is removed from each kinematic distribution. The resulting $34\times 34$ correlation matrix has a condition number of $2\cdot 10^2$.

\subsection{Combined Belle 23 and Belle II 23  experimental bins from HFLAV~\cite{HFLAV:2024}}
Combining experimental data sets is non-trivial. In fact, there are several sources of systematic uncertainties that are shared between different data sets and introduce correlations that have to be taken into account. Hence, we refrain from performing a naive combination of the Belle 23 and Belle II 23 data sets, because accounting for these correlations would require additional information that we do not possess. However, the Heavy Flavour Averaging Group (HFLAV) can account for them and provides combined results to be used by the wider community. \\
Therefore, for our analysis we use the upcoming HFLAV 24 \cite{HFLAV:2024} results. 
They are given  for the normalised differential decay rate in terms of 40 bins (each 10 bins in $w$, $\cos \theta_l$,  $\cos\theta_v$ and $\chi$). As above we remove the last bin in each channel. This reduces the condition number of the covariance matrix from $2\cdot 10^{18}$ down to $6\cdot 10^{2}$.

\section{Other input}
All the inputs that we employ in our analyses are in Tab.~\ref{tab:inputs}.
\begin{table}
\begin{center}
\begin{tabular}{l|c|l|l|llllll}
\hline\hline
observable&units&value&ref.&comment\\\hline
$M_B$           &GeV    &$5.2795$&\cite{Workman:2022ynf}&av. of $M_{B^+}$, $M_{B^0}$\\
$M_{D^\ast}$    &GeV    &$2.008555$&\cite{Workman:2022ynf}&av. of $M_{D^{*,+}}$, $M_{D^{\ast,0}}$\\
$M_e$           &GeV    &0.511&\cite{Workman:2022ynf}\\
$M_\mu$         &GeV    &0.10565&\cite{Workman:2022ynf}\\
$M_\tau$        &GeV    &1.77686&\cite{Workman:2022ynf}\\
\hline
\multirow{4}{*}{$M_{f,\Fone}$}       &
\multirow{4}{*}{GeV}    &6.739(13) &
\multirow{4}{*}{\cite{Dowdall:2012ab,Godfrey:2004ya}}&
\multirow{4}{*}{$1^+$ poles}\\
        &   &6.750&&\\
        &   &7.145&&\\
        &   &7.150&&\\
        \hline
\multirow{3}{*}{$M_\Ftwo$}       &
\multirow{3}{*}{GeV}    &6.275(1)&
\multirow{3}{*}{\cite{ParticleDataGroup:2016lqr,Godfrey:2004ya,McNeile:2012qf}}&
\multirow{3}{*}{$0^-$poles}\\
        &   & 6.842(6)&&\\
        &   & 7.250&&\\
        \hline
\multirow{3}{*}{$M_g$}       &
\multirow{3}{*}{GeV}    &6.329(3) &
\multirow{3}{*}{\cite{ParticleDataGroup:2016lqr,Dowdall:2012ab,Colquhoun:2015oha,Rai:2013xvr}}&
\multirow{3}{*}{$1^-$ poles}\\
        &   &6.920(18)&&\\
        &   &7.020&&\\
\hline
$\chi_{1^+}^{T}$  &GeV$^{-2}$&$3.894 \cdot 10^{-4}$&\cite{Bigi:2016mdz,Bigi:2017jbd,Bigi:2017njr}&for $f$ and $\Fone$\\
$\chi_{1^+}^{L}$  & 1        &$1.9421\cdot 10^{-2}$&\cite{Bigi:2016mdz,Bigi:2017jbd,Bigi:2017njr}&for $\Ftwo$\\
$\chi_{1^-}^{T}$  &GeV$^{-2}$&$5.131 \cdot 10^{-4}$&\cite{Bigi:2016mdz,Bigi:2017jbd,Bigi:2017njr}&for $g$\\
\hline
$\mathcal{B}(\BtoDstarellnu)$
                &1 &0.0497(12)&\cite{Workman:2022ynf}\\
$\tau(B^0)$     &$s$    &$1.519(4)\cdot 10^{-12}$      &\cite{Workman:2022ynf}\\
\hline
$\eta_{\rm EW}$ &1&1.0066(50)&\cite{Sirlin:1981ie}\\
\hline\hline
\end{tabular}
\caption{Summary of input.}\label{tab:inputs}
\end{center}
\end{table}
\bibliographystyle{jhep}
\bibliography{biblio}
\end{document}